\documentclass{article}
\usepackage{arxiv}

\usepackage[utf8]{inputenc} 
\usepackage[T1]{fontenc}    
\usepackage{hyperref}       
\usepackage{url}            
\usepackage{booktabs}       
\usepackage{amsfonts}       
\usepackage{nicefrac}       
\usepackage{microtype}      

\usepackage{graphicx}

\usepackage{esvect}
\usepackage{amsmath}
\usepackage[tight,footnotesize]{subfigure}
\usepackage{appendix}
\usepackage{lscape}
\usepackage{textcomp}
\usepackage{amssymb}
\usepackage{lscape}
\usepackage[noend]{algpseudocode}
\usepackage[boxruled]{algorithm2e}
\usepackage{hhline}
\usepackage{multirow}
\usepackage{lettrine}
\usepackage{tabularx}
\usepackage{hhline}
\usepackage{adjustbox}
\usepackage{caption}
\usepackage{makecell}
\usepackage{tabularx}
\usepackage{enumerate}

\title{A novel evolutionary-based neuro-fuzzy task scheduling approach to jointly optimize the main design challenges of heterogeneous MPSoCs}

\author{
  Athena Abdi \\
  Faculty of Computer Engineering\\
  K. N. Toosi University of Technology\\
  Tehran, Iran \\
  \texttt{a\_abdi@kntu.ac.ir} \\
\and
  Armin Salimi-Badr \\
  Faculty of Computer Science and Engineering\\
  Shahid Beheshti University\\
  Tehran, Iran \\
  \texttt{a\_salimibadr@sbu.ac.ir} \\
}

\begin{document}
\maketitle

\begin{abstract}
In this paper, an online task scheduling and mapping method based on a fuzzy neural network (FNN) learned by an evolutionary multi-objective algorithm (NSGA-II) to jointly optimize the main design challenges of heterogeneous MPSoCs is proposed.
In this approach, first, the FNN parameters are trained using an NSGA-II-based optimization engine by considering the main design challenges of MPSoCs including temperature, power consumption, failure rate, and execution time on a training dataset consisting of different application graphs of various sizes. Next, the trained FNN is employed as an online task scheduler to jointly optimize the main design challenges in heterogeneous MPSoCs. Due to the uncertainty in sensor measurements and the difference between computational models and reality, applying the fuzzy neural network is advantageous in online scheduling procedures.
The performance of the method is compared with some previous heuristic, meta-heuristic, and rule-based approaches in several experiments. Based on these experiments our proposed method outperforms the related studies in optimizing all design criteria. Its improvement over related heuristic and meta-heuristic approaches are estimated $10.58\%$ in temperature, $9.22\%$ in power consumption, $39.14\%$ in failure rate, and $12.06\%$ in execution time, averagely.  Moreover, considering the interpretable nature of the FNN, the frequently fired extracted fuzzy rules of the proposed approach are demonstrated.\\

\textbf{Keywords:}Task Scheduling, Heterogeneous MPSoC, Fuzzy Neural Network, Multi-objective Optimization, Interpretability.
\end{abstract}

\section{Introduction}
\label{sec:intro}

Along with technology advances, computer systems are widely used in various aspects of the life. This extensive demand makes their applications more complicated over the time. Nowadays, high performance computing is the most important and common requirement of all applications. These applications play a key role, in various aspects of human life such as industry, medicine, entertainment, financial structures, communications and so on~\cite{marwedel_2021embedded,henkel_2021dependable,yoo_2018low}. To deal with performance requirement, improving the frequency of computer processors is not accountable anymore due to its side effects on the other limitations as power consumption. Thus, employing parallelism approaches in processing such as integrating multiple processing core on a chip is widely used. Multiprocessor systems on chip (MPSoCs) provide high level of parallelism and are appropriate choice in designing modern computing systems. Integrating various processing units with different capabilities and characteristics on a chip, leads to heterogeneous MPSoCs~\cite{wolf2009multiprocessor,multiprocessor_book,soft_2020_whale}. Heterogeneous MPSoCs are more efficient in providing high performance platform along with controlling the other challenges such as power consumption and temperature. As a consequence, they are known as the de facto standard in designing embedded and cyber physical systems as the most applied computing systems in real life usages~\cite{ferrandi2010ant,akbari_2017_GA}.

Designing heterogeneous MPSoCs as an efficient and complicated system, comes with some challenges. Along with performance, other characteristics such as power consumption, chip temperature, lifetime reliability, cost and area of these systems should be considered and adjust during their design process~\cite{ERPOT,wolf2009multiprocessor,marwedel_2021embedded,HYSTERI}. Since these systems are widely employ in autonomous applications, their power budget is limited and generally dependent on batteries. Moreover, with technology advances and shrink in size of transistors, the leakage current that leads to static power and chip temperature increases. Increasing the chip temperature causes the hot spots and unavoidable physical damages on the chip~\cite{power_2021}. In addition, these events affect on the lifetime reliability of the system and make them more vulnerable against permanent faults that lead to system failures. Since modern computing systems are widely used in safety-critical applications to improve their precision and efficiency, decreasing the lifetime reliability is another design challenge of MPSoCs that should be considered during the design~\cite{chantem_hotspot,srinivasan2004case,chantem_2021}.

Dealing with the mentioned design challenges is possible at various levels of abstraction~\cite{ammar_2016performance,scheffer2018eda,mr_cross_2018}. System-level methods are very appropriate due to their flexibility, low overhead, and efficiency. Task scheduling and mapping is a system-level process that assigns the application tasks to various processing cores and orders their execution in time. The system requirements and constraints directly affect this process. Thus, optimizing the main design challenges of the system during task scheduling and mapping is the most effective system-level approach. This process is very complicated and known as an NP-hard problem~\cite{ammar_2016performance,henkel_2021dependable,ferrandi2010ant,marwedel_2021embedded}.
Consequently, various heuristic and meta-heuristic approaches are proposed to optimize some design challenges of MPSoCs during task scheduling. Based on the employed multi-objective optimization approach, these methods generate a unique solution or a set of points in form of Pareto front~\cite{ERPOT,sheikh_2016sixteen,chantem_hotspot,zhou_power}.
Task scheduling could be performed statically at design time or dynamically at runtime. Moreover, hybrid approaches are also presented that solve this problem at two phases~\cite{HYSTERI,mr_cross_2018,hybrid_2021}. Static approaches are capable of performing an exact exploration and are the appropriate choice for well-known applications where the execution structure of the system is determined. Otherwise, online approaches are required to adjust the optimization process based on the system conditions that are determined during its execution. Dynamic approaches enforce performance overhead to the system and their computation should be as low as possible~\cite{ammar_2016performance,chantem_2021_TCAD}. The existing complexity in modern applications and integration of devices on a single chip makes the uncertainty and inexact estimations more probable. Since static task scheduling approaches are incapable of handling this uncertainty and noisy measurement, employing online control mechanisms during the task scheduling process is essential.

Fuzzy inference systems are effective solutions to properly encounter uncertainty. Since these systems are built upon fuzzy logic, they can model uncertainty mathematically in the input variables considering them as linguistic variables \cite{Zadeh75,mendel2017uncertain, de2020fuzzy}. Therefore, they are successfully applied to applications encountering uncertainty like control, function approximation, resource allocation, and time-series prediction \cite{WU2021498,SALIMIBADR2022108258,SALIMIBADR2022139,LUO2019,Ebadzadeh2017,SalimiBadr2022,Ebadzadeh15,Salimi-Badr2017,ANFIS,SOFMLS,Das15,BAKLOUTI2018,ashrafi2020it2}. Considering the online scheduling as a control application that assigns different cores the ready task based on the received uncertain sensor measurements as the feedback, it is expected to successfully apply a fuzzy inference system to this problem. Moreover, another benefit of utilizing fuzzy inference systems is their interpretable "IF-THEN" rules which are expert-understandable.

Our proposed evolutionary-based neuro-fuzzy task scheduling approach aims at jointly optimizing the main design challenges of heterogeneous MPSoCs. It employs a fuzzy neural network (FNN) that is learned by NSGA-II to optimize system utilization, chip temperature, power consumption, and lifetime reliability. During this learning phase, the ruleset of the FNN is extracted statically based on exploring the various design solutions to optimize the target design challenges for several synthetic and real-life applications. Afterward, the learned FNN performs as the online and real-time scheduler on the target system that is capable of optimizing the main design challenges efficiently independent of the applications. Moreover, it efficiently handles the existing uncertainty and noises between the modeled and measured parameters by its fuzziness.
The main contributions of the proposed method are summarized as follows:

\begin{enumerate}
    \item Presenting an efficient online scheduling and mapping algorithm to assign tasks based on the current state of cores considering the degree of appropriateness;
    \item Presenting a Fuzzy Neural Network for implementing the proposed online task scheduling and mapping;
    \item Learning the parameters of the fuzzy rules using NSGA-II to jointly optimize the main challenges of heterogeneous MPSoCs;
    \item Extracting general application-independent and interpretable fuzzy rules for task scheduling and mapping process;
    \item Utilizing the improvement techniques including the dynamic voltage and frequency scaling (DVFS) and adding cooling slacks to efficiently explore the design space at the design-time (learning phase) and adapting the main parameters at the runtime.
\end{enumerate}

The rest of this article is organized as follows: First previous studies are reviewed in section \ref{sec:related}. Afterward, the necessary preliminaries including utilized application and architecture model, the scheduling and mapping, design challenges of MPSoCs, multi-objective optimization methods, and fuzzy inference systems are presented in section \ref{sec:pre}. Section \ref{sec:method} presents the proposed method including the problem statement, scheduling algorithm along with the architecture of the fuzzy neural network and learning method to realize it. Experimental results to evaluate the performance of the model along with comparisons with the previously proposed methods are presented in section \ref{sec:results}. Finally, section \ref{sec:concusion} presents the conclusions.

\section{Related work}
\label{sec:related}

Task scheduling and mapping problem is one of the most effective system-level approaches in optimizing the main design challenges of heterogeneous MPSoCs. Since the complexity of this problem is known as NP-hard, several heuristic and meta-heuristic approaches are presented to dealing with it~\cite{ferrandi2010ant,soft_2020_whale,softcom_ant_2018}. These approaches are focused on various design challenges as objectives. Most of them are concentrated on system performance along with one of the other system parameters like lifetime reliability, power consumption, or chip temperature. Moreover, these approaches could be performed statically or dynamically that mitigate the problem at design or runtime~\cite{das_2018literature,ferrandi2010ant,MEJ}.

Heuristic approaches try to optimize the design challenges through an efficient and lightweight approach. They aim at various design challenges and solve the problem during the design or runtime. In~\cite{power_2008}, two heuristic approaches based on the system gradient are proposed that minimize the energy consumption and performance of heterogeneous MPSoCs. These iterative algorithms start task scheduling and mapping with a greedy performance-feasible solution and then by considering the gradient of energy consumption, optimize it gradually by utilizing available voltage and frequency levels of processing cores. In~\cite{power_2016}, the leakage power is aimed and a set of heuristic solutions to optimize it along with system performance is presented. In these approaches, the system temperature as the main effective system-level parameter on static power is studied and tries to minimize it at the core and system level. In~\cite{sheikh_2016sixteen} the joint optimization of performance, power consumption, and temperature is aimed heuristic scheduling approaches are presented. These methods are mainly based on the integration of objectives by a weighted sum that leads to one optimal point in the multi-objective optimization process. Here, adjusting the weights based on the range of parameters and their importance is critical and directly affects the appropriateness of the final result. These parameters could be various in different applications so exploring the whole design space and presenting a solution set is more effective.
In~\cite{Ekhtiyari2019}, a task scheduler based on a fuzzy inference system is presented to optimize power consumption and temperature simultaneously. In this approach, the static and predefined fuzzy rules are employed to determine the appropriate processing core and the scheduling details are defined based on improving the system temperature and utilization.

Meta-heuristic approaches try to explore the solution space and optimize the design challenges during the task scheduling process more efficiently. Genetic algorithm is one of the most effective meta-heuristic algorithms that is compatible with task scheduling due to its operators~\cite{akbari_2017_GA,xu2014_genetic}. In~\cite{das_2014combined} a task scheduling approach based on NSGA-II to jointly optimize performance, power consumption, and the soft error rate of MPSoCs is presented. This approach, explores the design space for appropriate scheduling based on the static application model of the system to optimize performance, power consumption, and soft error rate. In~\cite{MEJ} the optimization target is expanded and a static task scheduling approach to optimize performance, lifetime reliability, power consumption, and temperature of heterogeneous MPSoC is proposed. Here the design space is explored by an NSGA-II-based engine and the various failure rate mechanisms are considered separately during the optimization process. Since these approaches explore the design space more comprehensively than heuristic methods, their process time is very high. But due to the appliance of these approaches in design time, it is tolerable in well-known applications that the application and architecture models are determined.

Today due to the various applications of MPSoCs, considering static scheduling based on a predefined scenario is not enough. Employing hybrid or dynamic scheduling approaches that are adaptive to the system application and its state is more effective~\cite{HYSTERI,mr_cross_2018,das_2018literature}.
In~\cite{HYSTERI}, a hybrid task scheduling approach based on the list scheduling approach is presented to jointly optimize performance, lifetime reliability, power consumption, and chip temperature. This method generates a solution set based on Pareto optimization and then performs a fine-tuning based on a lightweight static online temperature controller.
Enhancing the online control of hybrid approaches, make them more efficient in optimizing the main design challenges during task scheduling. Since the complexity of this dynamic control affects directly system performance, its overhead and efficiency should be studied in detail.

\section{Preliminaries}
\label{sec:pre}

\subsection{Application and Architecture Models}
\label{sec:apparch}
\begin{figure}
\centering
\begin{tabular}{cc}
    \subfigure[][]{\includegraphics[width=1.7in,height=2in]{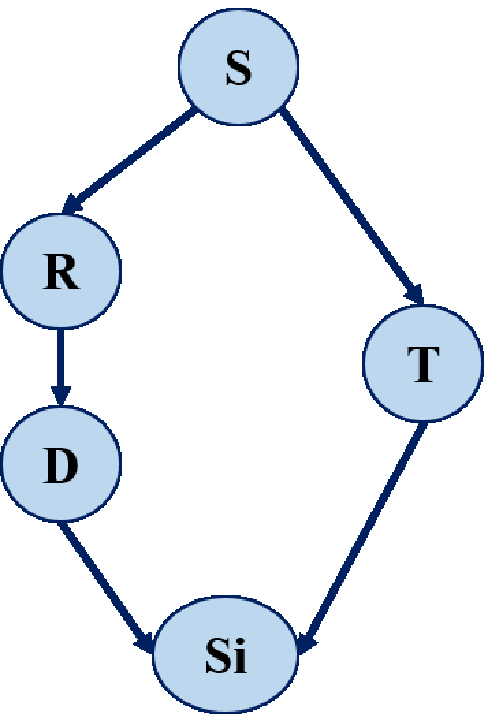}} & \qquad
    \subfigure[][]{\includegraphics[width = 2in]{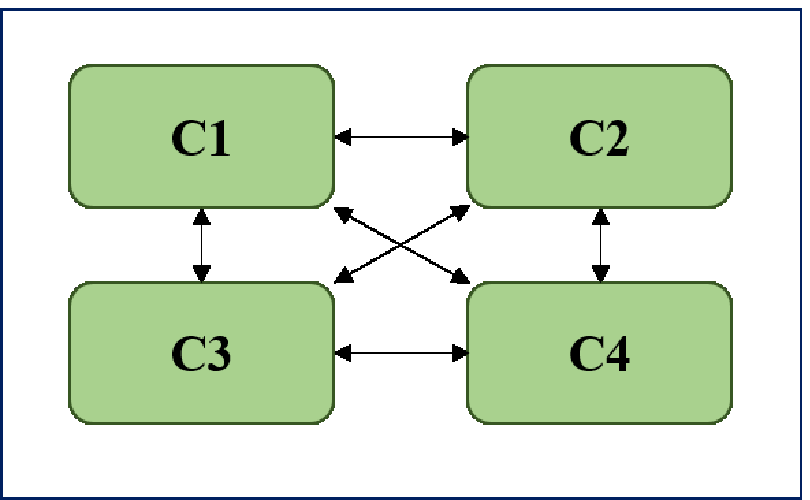}}
\end{tabular}
\caption{(a)Sample application graph representation in form of the DAG of tasks, (b) Sample architecture graph of a heterogeneous multiprocessor system.}
\label{fig:app-arch}
\end{figure}

We consider the application as a directed acyclic graph (DAG) that consists of independent tasks as nodes and their data dependency as edges. This $\mathcal{APP}$ graph, is formally defined as a 5-tuple: $(V_{app},E_{app},W_{app},D_{app},H_{app})$. Where these tuples represent the tasks, their data dependency relation, cost of communication, worst case execution time and heterogeneity property, respectively. The data dependency relation of the tasks is defined as the predecessor-successor relation and determines based on the application. This data dependency and communication has a cost that is represented as the weight of the edges in $\mathcal{APP}$ graph. Moreover, the worst case execution time of each application task is determined based on the nominal frequency of the system and the heterogeneity shows the compatibility and execution requirement of each task on heterogeneous platforms. Figure~\ref{fig:app-arch}-a shows an example of the defined application model for Office application of E3S benchmark suits~\cite{E3S}.

Our considered architecture is a heterogeneous multi-core system on chip that is modeled as a graph. In this $\mathcal{ARCH}$ graph, the nodes are heterogeneous processing cores and the edges are their communication links. The heterogeneity of processing cores are defined in their execution capability and multiple levels of operational voltage and frequency levels are considered for each of them. Moreover, it is assumed that the processing cores are fully connected. Figure~\ref{fig:app-arch}-b shows a sample architecture graph of our assumed MPSoC platform.

\subsection{Scheduling and mapping}
\label{sec:scheduling}
Task scheduling and mapping problem in heterogeneous multiprocessor systems defines as determining the appropriate time and processing core from architecture graph for executing each task of application graph. Making this system-level decision directly affects the main design parameters of the system such as performance, power consumption and lifetime reliability. During the mapping phase, the appropriate processing core for executing the current task is selected. This selection is tightly dependent on the accordance of task to the properties of processing core in heterogeneous system. Afterward, the execution of the task on the selected core is scheduled.

During these phases, various choices based on design space exploration are possible that lead to different performance, power consumption and lifetime reliability. Thus, dealing with this problem to optimizing the main design challenges of heterogeneous MPSoC at system level that provides low cost and high flexibility in design~\cite{wolf2009multiprocessor,abdallah2017advanced}. Moreover, task scheduling and mapping process could performed statically at design time or dynamically at runtime. Hybrid approaches aim at distributing the tasks during design and runtime in order to integrate their benefits. To this aim, we form a multi-objective optimization problem to optimize performance, power consumption, lifetime reliability and chip temperature during a hybrid task mapping and scheduling process.

\subsection{Design Challenges of Multiprocessor systems on chip}
\label{sec:objectives}
Integrating several processing capabilities in multiprocessor systems on chip (MPSoC), complicates their design process. The most important design challenges are performance, power consumption, reliability, chip temperature, cost and area. These parameters are not independent and have antagonistic relations that make their simultaneous mitigation during design more complicated~\cite{multiprocessor_book,ammar_2016performance,ERPOT,MEJ}.

\subsubsection{Lifetime Reliability}
Multiprocessor systems are vulnerable against transient and permanent faults that occur in processing units and commutations. Transient faults have temporal effect on the system while permanent ones affect the system lifetime.
Lifetime reliability of multiprocessor system is threaten through several failure mechanisms that have mainly intrinsic sources.These failure mechanisms could be divided to three classes:
\begin{itemize}
    \item Electro-migration (EM) and Stress-migration (SM) on the communication links,
    \item Time-dependent dielectric breakdown (TDDB) in gate oxide,
    \item Negative Bias Temperature Instability (NBTI) in transistors.
\end{itemize}

Electro-migration occurs when electrical current movement in the interconnections leads to movement of metal atoms. It mainly depends on current density and metal quality and its high-level empirical MTTF relation is derived as follows~\cite{failure_2009esd,JEP122H}:

\begin{equation}
        MTTF_{EM} = A_{EM} * (J - J_{circ})^{(-n)} * exp \quad (\frac{E_a}{KT})
\end{equation}
where, $A_{EM}$ is a scale factor, $J$ is the applied current density of interconnect, $J_{circ}$ is the threshold current
density, $n$ is the current density exponent, $E_a$ is the activation energy, K is Boltzmann's constant and T represents the temperature in Kelvins. It should be noted that in this model, J should be much greater than $J_{circ}$ to cause failure.

Stress migration occurs due to the atom movement in metal interconnections. It is mostly related to the thermal incompatibility of metal and dielectric. The same as EM, this mechanism lead to the failure in metal lines or adjacent contacts. The MTTF relation of this failure is computed as follows~\cite{failure_2009esd,JEP122H}:

\begin{equation}
        MTTF_{SM} = A_{SM} * (T_0 - T)^{(-n)} * exp \quad (\frac{E_a}{KT})
    \end{equation}
where, $A_{SM}$ is the scale factor, $T_0$ is stress-free temperature for metal, $T$ is temperature in Kelvins, $n$ is a material-dependent constant, $E_a$ is the activation energy and K is the Boltzmann's constant.

Time dependent dielectric breakdown (TDDB) occurs due to the gradual degradation of the gate dielectric. In this case, aggregation of electrical current in the gate oxide forms an electrical field and leads to the dielectric breakdown. The MTTF relation of this mechanism is computed by the following relation~\cite{failure_2009esd,JEP122H}:

    \begin{equation}
        MTTF_{TDDB} = A_{TDDB} * exp (-\gamma E_{ox}) * exp (\frac{E_a}{KT})
    \end{equation}
where, $A_{TDDB}$ is a scale factor, $\gamma$ is field acceleration parameter, $E_{ox}$ is the electric field across the dielectric, $E_a$ is the activation energy, K is Boltzmann's constant and T is the temperature in Kelvins.

 Negative bias temperature instability (NBTI) is the most effective factor in wear-out that leads to increasing the threshold voltage of transistors~\cite{nbti_2021bias,franco_2014reliability}. Changes in threshold voltage causes timing violations and failure of the system. The MTTF relation of this mechanism is computed as \cite{failure_2009esd,JEP122H}.

    \begin{equation}
        MTTF_{NBTI} = A_NBTI * (V_{GS})^{-\gamma} * exp (\frac{E_a}{KT})
    \end{equation}
where, $A_{NBTI}$ is a scale factor, $V_{GS}$ is the absolute value of the gate voltage applied to the PMOS in inversion, $\gamma$ is the voltage acceleration factor, $E_a$ is the activation energy, $K$ is the Boltzmann's constant and $T$ is the temperature in Kelvins.

In our proposed scheduling approach, to consider the effect of the explained failure mechanisms simultaneously, sum-of-failure-rates (SOFR) measure~\cite{srinivasan2004case} is employed. Moreover, temperature is the most effective and common term in all described failure rates. Thus to estimate the failure rate of our assumed MPSoC, the mean time to failure (MTTF) of each processor by applying SOFR measure and its temperature is considered. The MTTF based on SOFR measure is computed as follows~\cite{srinivasan2004case}:

\begin{equation}
\label{eq:MTTF_SOFR}
MTTF = \displaystyle \frac {1}{\lambda_{total}} = \displaystyle \frac{1}{\displaystyle \Sigma_{p=1}^j \displaystyle \Sigma_{l=1}^k \lambda_{pl}}
;
\end{equation}
where $\lambda_{pl}$ represents the failure rate of $p^{th}$ processing unit because of the $l^{th}$ failure mechanism.

In this paper to taking the lifetime reliability into account, the failure rate is considered. Since failure rate is related to time and is not invariant of the number of scheduled tasks, here we utilize the global system failure rate (GSFR) measure instead of failure rate~\cite{assayad_2013tradeoff,alain_2009,ERPOT}. The GSFR of a schedule that consists of application tasks is computed as follows:

\begin{equation}
\label{eq:gsfr}
    \Lambda (S) = \displaystyle \frac{\lambda (S)}{\sum_{(\tau_i,c_i,f_i)\in S} \mathcal{E}xe (\tau_i,c_i,f_i)}
\end{equation}
here, S is the scheduling of application tasks on processing cores, $\lambda$ is the failure rate of scheduling based on the described SOFR and $\mathcal{E}xe$ is the execution time of the tasks of application on the processing cores of underlying architecture.

\subsubsection{Power Consumption}
Along with increase in processing capability of MPSoCs, their power consumption is growing dramatically. Moreover, these systems are widely used in autonomous embedded applications which are battery-dependent and have limitation in energy resource. Thus, considering power consumption during MPSoC design is required.
Power consumption of executing a task on a processing core consists of two aspects: (i) static and (ii) dynamic~\cite{zhou_power,power_2021}.
Dynamic power is dependent on the operational voltage and frequency and the static part is mainly related to leakage power and system temperature. The overall power consumption of MPSoCs is derived by summing up the dynamic and static parts that are computed in system level as follows:

\begin{equation}
\label{eq:power}
    P_{system} = P_{dynamic} + P_{leakage} = C_{eff} \displaystyle \times V ^2 \displaystyle \times f + \alpha \displaystyle \times T(t) + \beta ;
\end{equation}
where $C_{eff}$ represents the switching capacitance, V and f are operational voltage and frequency, respectively.
Moreover, $\alpha$ and $\beta$ are architecture-dependent coefficients that are determined based on the characteristics of the platform~\cite{thiele_2013,lee_2021thermal} and T represents the chip temperature in \textit{Kelvin}.

\subsubsection{Chip Temperature}
Along with technology advances, increase in power density and rate of permanent failures, the chip temperature has become important. The instantaneous temperature of a MPSoC depends on its power consumption and the current temperature. Based on the Fourier
law of heating using an RC equivalent thermal model, the temperature is computed as follows~\cite{chantem_2011,crc_thermal}:

\begin{equation}
\label{eq:theta}
  C \left( \displaystyle \frac{dT(t)}{dt} \right) + G \big( T(t) - T_{amb} \big) = P(t) ;
\end{equation}
where $C$ and $G$ are thermal capacitance and conductance receptively, $t$ is time, $T(t)$ represents instantaneous temperature and $T_{amb}$ is the ambient temperature. $P$ expresses the power consumption of the system based on the dynamic and static aspects as described in Eq.~(\ref{eq:power}).

Due to the shrinking size of chips and high integration rate in the modern technology, the effect of temperature of the neighbors on each core becomes important. To consider this effect, in this paper we use the following two-dimension spatial heat transfer equation among processing cores~\cite{chantem_2011,CHENG_temp_2021}:

\begin{equation}
  \mathit{Heat\_transfer_{2D}} = \!\! \displaystyle \sum_{c' \displaystyle \in nbr(c)} \!\!
  G(c,c') \big( T_c(t) - T_{c'}(t) \big)
  \label{eq:Temp-2D}
\end{equation}
where $nbr(c)$ is the neighbor set of core~$c$, $T_c$ and $T_{c'}$~are the temperatures of core~$c$ and its neighbor~$c'$, $G(c,c')$ is the thermal conductance between cores $c$ and~$c'$ which depends on their distance and chip geometry characteristics.

By applying the power consumption as Eq.~(\ref{eq:power}) and heat transfer between adjacent cores as Eq.~(\ref{eq:Temp-2D}), the temperature of each processing core can be expressed as:

\begin{equation}
 \begin{array}{ll}
    C \left( \displaystyle \frac{d T_c(t)}{dt} \right)  = & - G \big( T_c(t) - T_{amb} \big) - \displaystyle \sum_{c' \displaystyle \in nbr(c)} G(c,c') \big( T_c(t) - T_{c'}(t) \big) \\
    & + C_{eff} V^2 \cdot f + \alpha \cdot T_c(t) +\beta
 \end{array}
\label{eq:theta_rewrite}
\end{equation}

To solve this differential equation, we have:
\begin{equation}
    \begin{array}{ll}

        \displaystyle \frac{dT_c(t)}{dt} = -a \cdot T_c + b, &  with \\

         & \\

        a = \displaystyle \frac{G - \alpha + \sum_{c' \in nbr(c)} G(c,c')}{C}
        \nonumber & \\

         & \\

        b = \displaystyle \frac{G \cdot T_{amb} + \!\! \displaystyle \sum_{c' \in nbr(c)} \!\!
        T_{c'}(t) \cdot G(c,c') + C_{eff} \cdot V^2 \cdot f + \beta}{C}
    \end{array}
\end{equation}
Thus, the closed form solution of the temperature equation is defined as follows:

\begin{equation}
\label{eq:theta-total}
    \begin{array}{lr}
        T_c(t) = T^{\infty} + \big( T(t_0) - T^{\infty} \big) e^{-a(t-t_0)} & with \\
         & \\
        T^{\infty} = \frac{b}{a} &
    \end{array}
\end{equation}
where $T(t_0)$ and $T^{\infty}$ are the initial and  steady state temperatures, respectively~\cite{chantem_2011,crc_thermal}

\subsection{Multi-objective optimization methods}
\label{sec:multi}
\begin{figure}
    \centering
    \includegraphics[width=2.7in]{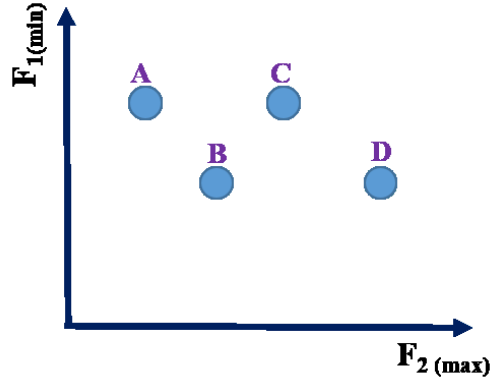}
    \caption{Domination example in two-dimensional space (F1,F2)}
    \label{fig:paretodom}
\end{figure}

multi-objective optimization, aims at optimizing two or more conflicted criteria during a decision making problem. In this context, simultaneous consideration of all objectives and meet their contradicted constraints is very complicated. Moreover, dealing with non-comparable solutions during the optimization process is also challenging. Due to existing trade-off among objectives and their contradictions, the final solution is not unique. For instance, minimizing power consumption and maximizing the performance of a chip are not aligned. Thus, selecting the best performance greedily leads to losing power consumption and vice versa.
Pareto optimization is an appropriate approach in facing the mentioned challenges of multi-objective optimization. It explores the whole solution space and utilizes the dominance rule to compare the generated solutions and meet their existing trade-off. Dominance is used to determine the appropriateness of a solution over others. In the two-dimensional space $(X_1,X_2)$ $X_1$ dominates $X_2$ if:
\begin{itemize}
    \item $X_1$ is better than $X_2$ in at least one objective,
    \item $X_1$ is not worse than $X_2$ in all objective.
\end{itemize}
Figure~\ref{fig:paretodom}, shows the domination in two-dimensional space $(F_1,F_2)$. Here, "B" and "C" dominate "A" and are dominated by "D" but they do not dominate each other.
The set of dominance points in a solution space construct the "Pareto Front".

Constructing the whole Pareto front and simultaneous consideration of all objectives are not trivial. To this aim various approaches such as \textit{aggregating}, \textit{hierarchization}, transformation and population-based methods are utilized~\cite{ERPOT,scheduling_book_2006}. The three former approaches convert the problem to the single objective form by integrating and prioritizing the objectives or transforming all but one to the constraints. These approaches especially the aggregation and hierarchization suffer from inaccuracy. Since they are dependent on application and do not cover the whole Pareto front. Transformation ($\epsilon$-Constraint method) is more effective in building the Pareto front but determining the proper step ($\epsilon$ value) in iterative exploration of the design space is challenging~\cite{ERPOT}.
Considering population-based approaches such as genetic algorithm and ant colony leads to exploring the whole design space and building the Pareto front more comprehensively.

\subsubsection{Non-dominated Sorting Genetic Algorithm}

Non-dominated sorting genetic algorithm (NSGA-II) is one of the most proper approaches in exploring the design space and building the Pareto front. It classifies the solutions into some non-dominated sets that called fronts. Then it ranks the initial random solutions based on their covering front and determines the crowding distance factor for them.
The rank of each front is assigned based on the domination level. All non-dominated solutions are assigned to the first front and the ones that are dominated by the first level solutions located to the second front and so on~\cite{nsga}.

\begin{figure}[t]
    \centering
    \includegraphics[width=5.5in]{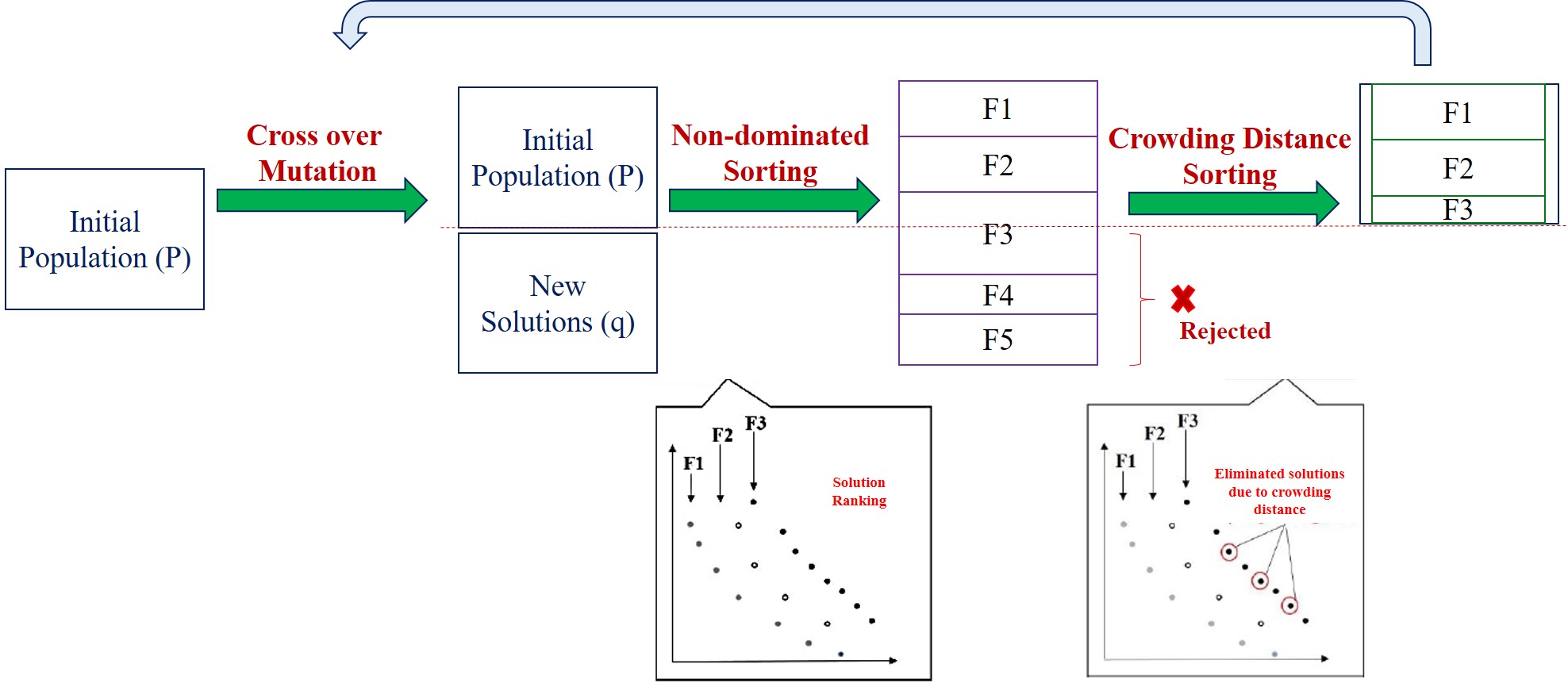}
    \caption{Ranking and selection of solutions based on dominance and crowding distance measures in NSGA-II algorithm}
    \label{fig:NSGA}
\end{figure}

Based on the population size in each front, this approach evaluates the solutions based on their rank and crowding distance. Crowding distance guarantees the diversity of the solutions and avoid their concentration and biasness.
At each generation, solutions with better ranks and crowding distance are remained and the others eliminate.
Then, to make more diversity in the population and explore the solution space, cross-over and mutation operators are applied to the population. The new and old solutions are ranked again and based on the population size the most eligible ones are kept. The explained process are repeated on the new population and iterates until the convergence is met~\cite{nsga}.
Figure~\ref{fig:NSGA}, shows the explained mechanism and the selection and ranking processes. In this figure, $F_i$ is the hierarchy of the fronts and the domination levels.

\subsection{Fuzzy Inference Systems}
\label{sec:fis}
\begin{figure}[h]
    \centering
    \includegraphics[width=5in]{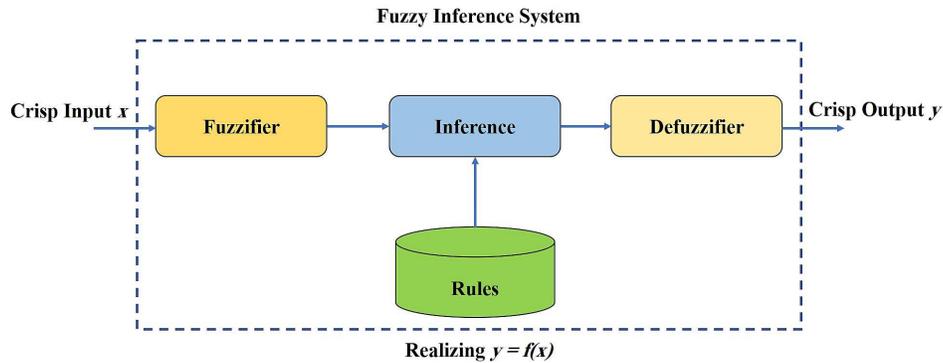}
    \caption{Fuzzy Inference Systems structure. This structure receives a crisp input \textit{x}, fuzzifies it through \textit{Fuzzifier} component, infers proper fuzzy output through \textit{Inference} component based on its \textit{rules}, and finally produces the crisp output using the \textit{Defuzzifier} component.}
    \label{fig:fis}
\end{figure}

Fuzzy Inference Systems (FISs) are rule based decision making systems including a set of "IF-THEN" linguistic rules \cite{SALIMIBADR2022108258,SALIMIBADR2022139,LUO2019,de2020fuzzy,Ebadzadeh2017,SalimiBadr2022,Ebadzadeh15,Salimi-Badr2017,ANFIS,SOFMLS,Das15,BAKLOUTI2018,ashrafi2020it2}. One of the main advantage of these systems is their interpretability. Since these systems are built upon using linguistic variables based on fuzzy linguistic values, their decisions can be understood by experts. Moreover FIS considers the uncertainty which is helpful to face noisy measurements and imprecise computational models which are inevitable in applications like embedded and cyber-physical systems \cite{marwedel_2021embedded,HYSTERI}. Figure \ref{fig:fis} shows different components of a FIS \cite{mendel2017uncertain}.

Generally the form of $i^{th}$ \textit{Zadeh}'s fuzzy rule of a FIS is as follows \cite{mendel2017uncertain}:

\begin{equation}\label{eq:rule}
  \text{$R^{i}$: IF $x_1$ is $A_1^i$ and $x_2$ is $A_2^i$ and ... and $x_n$ is $A_n^i$  THEN $y$ is $y^i$}
\end{equation}
where $x_j$ ($j = 1,2,...,n$) is the $j^{th}$ input variable, $A_j^i$ is the fuzzy set representing a lingual value for $j^{th}$ variable of the $i^{th}$ fuzzy rule, $y$ is the output value, and $y^{i}$ is the output value proposed by the $i^{th}$ fuzzy rule.

To add learning ability to FISs, Fuzzy Neural Networks (FNN) are proposed that have the interpretability advantage of FIS along with adaptive structure of Neural Networks \cite{LUO2019,de2020fuzzy,ANFIS,SALIMIBADR2022108258}.

In the proposed method, a Fuzzy Neural Network is used to determine the assignment degree of a core to accept a task in the scheduling process based on its state.

\section{Proposed method}
\label{sec:method}
This section presents the details of our target problem and the proposed task scheduling approach.

\subsection{Problem statement}
\label{sec:state}
The objective of our proposed task scheduling approach is to optimize the main design challenges of heterogeneous MPSoCs including: utilization, lifetime reliability, power consumption and temperature. To this aim, a fuzzy neural network learnt by an evolutionary multi-objective optimization algorithm (NSGA-II) is employed. Our assumptions and objectives could be described as follows:

\textbf{Given:}
\begin{itemize}
    \item The target heterogeneous MPSoC modeled as an architecture graph of processing cores and their interconnection,
    \item The set of operational voltage and frequency levels for each processing core,
    \item The application modeled as a weighted directed acyclic graph composed of tasks and their precedence relations,
    \item The worst case execution time (WCET) of executing each application task on the various processing cores of architecture graph based on their nominal frequency,
    \item The variation range of lifetime reliability, power consumption and temperature based on the characteristics of architecture platform.
\end{itemize}

\textbf{Goals:}
\begin{itemize}
    \item Online scheduling and mapping of application tasks on heterogeneous processing cores though a fuzzy neural network approach such that:
    \begin{itemize}
        \item Optimize the main design challenges of heterogeneous MPSoCs including: system utilization, lifetime reliability, power consumption and temperature jointly,
        \item Meeting the precedence relation of application tasks,
        \item Distributing tasks on heterogeneous processing cores in order to balance their wear out rate,
        \item Mitigating the existing measurement gap between the static models and real sensors by employing fuzzy neural network as a decision maker,
        \item Determining a proper set of interpretable fuzzy rules to meet the existing trade-off among the defined optimization problem objectives.
    \end{itemize}
\end{itemize}

\subsection{Scheduling algorithm}
\label{sec:alg}
Optimizing the main design challenges of heterogeneous MPSoCs is considered at various abstraction levels. In this context, optimizing the design challenges during task scheduling and mapping as a system-level approach is very effective. This process could be performed statically at design time or dynamically at runtime.~\cite{das_2018literature,ammar_2016performance,scheduling_book_2006}.
In this paper an online task scheduler for heterogeneous MPSoCs to jointly optimize utilization, lifetime reliability, power consumption and chip temperature using fuzzy neural network is presented. The proposed scheduler starts its work along with the application execution but it has also a pre-process phase to learn the fuzzy neural network.

Since the tasks of application have precedence relation, it is required to determine the candidate tasks at each scheduling stage. The data dependency of candidate tasks are met and they are ready to execute. Based on the limitations of our underlying architecture and the available processing cores, a specific number of ready tasks are selected and executed in parallel. To sort the list of ready tasks at each scheduling stage, we define an emergency parameter that is related to the deadline and worst case execution time of each task. This emergency parameter demonstrates the the flexibility of each task and its postpone tolerance. Thus the priority of ready tasks at each scheduling step is determined as follows:

\begin{equation}
    urgent\_task (n) = argmin_{\tau \in Ready^{(n)}} (D_{\tau} - WCET_{\tau})
\end{equation}
where, n is the scheduling step and $\tau$ demonstrates the ready tasks. D and WCET are the deadline and worst case execution time of each ready task, respectively.

After constructing the ready list and sort it based on the defined emergency factor, it is required to determine appropriate processing cores for their execution. Moreover, there exist multiple levels of operational voltage and frequency on each processing core and the most proper one should be selected during the mapping phase.
At each scheduling step, based on the available processing core and the limitation of the underlying heterogeneous MPSoC a specific number of ready tasks are assigned. The task mapping decision is made for each ready task and based on optimize the main objectives of our target problem. Selecting the best execution option among the available processing core and frequency levels, is a multi-objective optimization problem and we solve it with a learned fuzzy neural network.

This fuzzy neural network is learned by NSGA-II as one of the most effective multi-objective optimization algorithms to jointly optimize the main design challenges of MPSoC during the task scheduling process. Fuzzy neural networks are interpretable and modifiable so their rules and parameters are flexible to the expert knowledge and system updates. Moreover, the learning process is performed based on several static cases that utilize the models of main system parameters. In fact these models are not exact and there exist difference between the estimated and actual values of main design parameters. To cover this variance that leads to uncertainty in decision, fuzzy neural network is an appropriate option.

The ready list of tasks, the available processing cores and their operational voltage and frequency levels are the input of the fuzzy neural network. It determines a rank for each possible mapping of ready task based on the joint optimization of the main objectives of our target problem. The mapping decision with lowest rank is the best in meeting the existing trade-offs among the objectives.
This rank determines the appropriateness of each processing core and its corresponding voltage/frequency level in executing the current ready task in terms of optimizing system utilization, lifetime reliability, power consumption and chip temperature.

Based on considering several application graphs, the explained process is repeated and the appropriate parameters of fuzzy neural network is learned through an NSGA-II-based approach.
To this aim, the joint optimization of considered design challenges is examined as the cost function and the parameters of fuzzy neural network is set based on several application graphs. These parameters are utilized during the task scheduling and mapping process of any application at runtime. The learning process makes our our proposed scheduler appropriate to jointly optimize the main objectives during system operation and for various applications.
The details of our proposed task scheduling approach is summarized in Algorithm~\ref{alg:scheduling}.
As well, the details of our proposed learning approach and the utilized fuzzy neural network architecture are explained in sections~\ref{sec:learning} and~\ref{sec:fnn}.

\begin{algorithm}[!t]
\scriptsize
  \textbf{Inputs:} \\
  $\quad$ \text{Application graph}\\
  $\quad$ \text{Architecture graph of underlying heterogeneous MPSoC} \\
  $\quad$ \text{Operational range of lifetime reliability, power consumption and temperature}\\
  \textbf{Output:}\\
   $\quad\,$ \text{A task scheduling and mapping to jointly optimize:}\\ $\quad\,$ \text{system utilization, lifetime reliability, power consumption and chip temperature (schedule)}

\caption{The proposed learning-based fuzzy task scheduling approach }
\label{alg:scheduling}
    \vspace {2mm}
    ready$\_$list $\gets$ tasks of application graph with satisfied data dependency condition\\

    \While {ready$\_$list $\neq$ $\emptyset$}{

    $\tau_{urgent} = argmin_{t \in ready\_list}
            (D_{t} - WCET_{t})$

    \For{$c \in available processing cores set and f \in operational voltage/frequency levels$}{
        - Compute: \\
        ($u(\tau_{urgent},c,f), \Lambda (\tau_{urgent},c,f), P(\tau_{urgent},c,f), \theta (\tau_{urgent},c,f)$),

        Degree(c,f) $\gets$ FNN ($u(\tau_{urgent},c,f), \Lambda (\tau_{urgent},c,f), P(\tau_{urgent},c,f), \theta (\tau_{urgent},c,f)$) \Comment{the details of FNN are presented in Algorithm~\ref{alg:FNN}}
    }

    - Determine the appropriate core along with operational frequency level: \\
    Selected$\_$Processing$\_$Unit $\gets argmin_{c,f}$ Degree

    schedule[n] $\gets (\tau_{urgent},Selected\_Processing\_Unit)$

    ready$\_$list $\gets$ tasks of application graph with satisfied data dependency condition

    n $\gets$ n+1
    }
\end{algorithm}

\begin{figure}[h]
    \centering
    \includegraphics[width=5.3in]{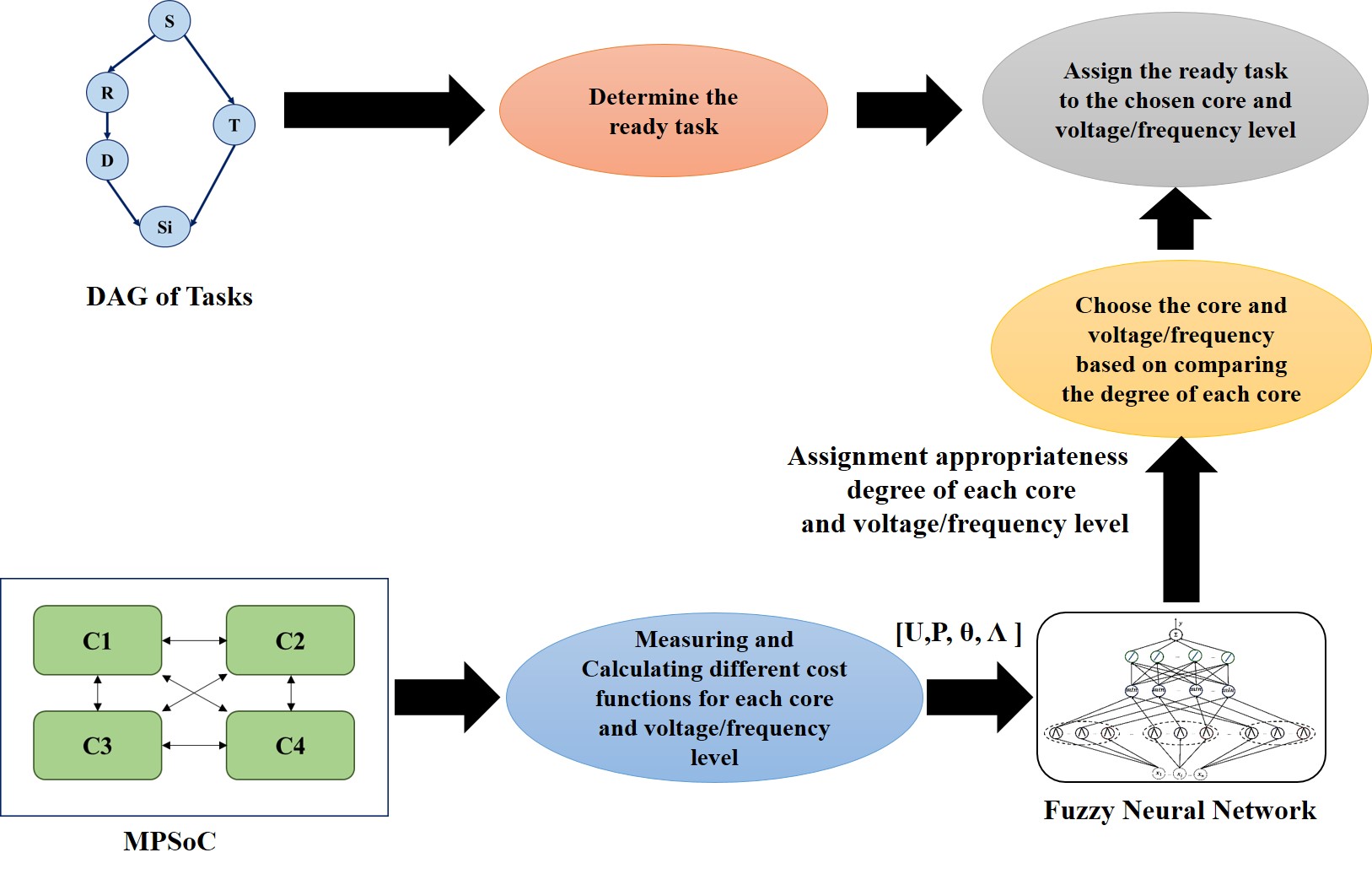}
    \caption{The work-flow of scheduling and mapping of a task. In this figure $\theta$ stands for a vector consists of the measured temperature values of different cores, $P$ is a vector consists of the calculated power values of different cores, $U$ is a vector consists of calculated utilization values of each core, and $\Lambda$ is a vector that its elements are the calculated GSFR values of different cores.}
    \label{fig:scheduling}
\end{figure}

The described process is repeated iteratively for various ready tasks during the task scheduling process. Along with task scheduling, the values of main parameters of underlying architecture is updating and compares to the system limitations.
Figure~\ref{fig:scheduling} shows the workflow of our proposed task scheduling and mapping approach.

\subsection{Fuzzy Neural Network}
\label{sec:fnn}

In the proposed method a fuzzy neural network is presented to determine the assignment degree of each core based on its current state. Figure \ref{fig:fnn} shows the architecture of this fuzzy neural network. This fuzzy neural network consists of five layers: 1- input layer, 2- fuzzification layer, 3- fuzzy rules layer, 4- normalization layer, and 5- output layer. The description of each layer is outlined as follows:

\begin{figure}[h]
    \centering
    \includegraphics[width=6in]{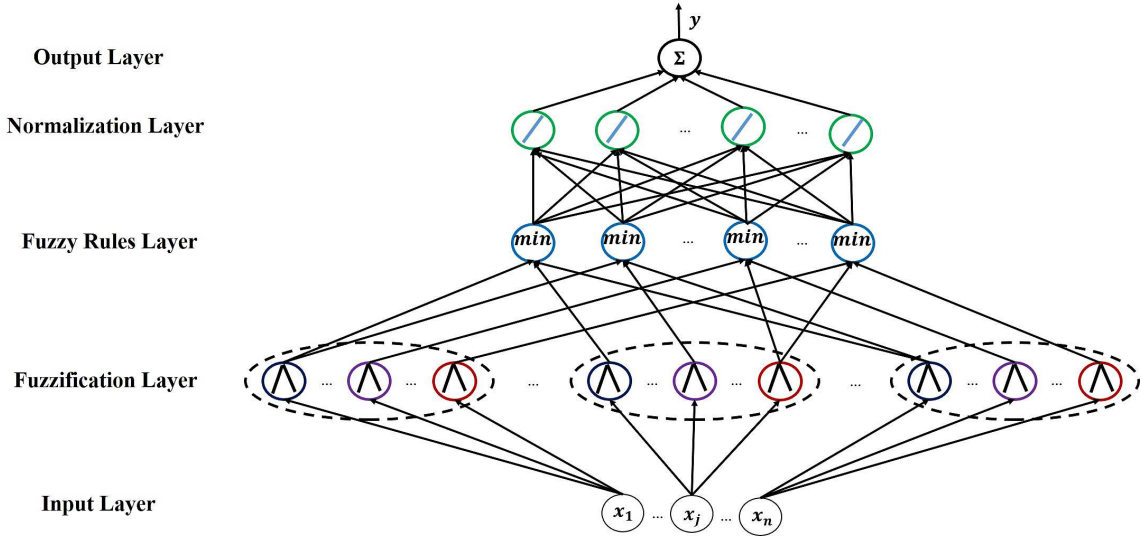}
    \caption{Fuzzy Neural Network architecture.}
    \label{fig:fnn}
\end{figure}

\begin{itemize}
\item{\textbf{Input Layer}}: This layer is composed of four neurons that deliver the variables describing the state of each voltage-frequency level of each core ($x_1=$utilization ($U$), $x_2=$power ($P$), $x_3=$temperature ($\theta$), and $x_4=\Lambda$ (global system failure rate ($GSFR$))).

\item{\textbf{Fuzzification Layer}}: Each neuron of this layer receives an input variable from the input layer and calculates its membership degree to a lingual value based on a triangular membership function. The membership value of $j^{th}$ input variable $x_j$ ($j=1,2,3,4)$ to the $l^{th}$ fuzzy set $A_{l}$ is calculated as follows:

\begin{equation}
    \mu_{A_{l}}(x_j) = \left\{\begin{array}{ll}
    \frac{(x_j-a_l)}{(c_l-a_l)} & a_l \leq x_j \leq c_l\\ \\
    \frac{(b_l-x_j)}{(b_l-c_l)} & c_l\leq x_j \leq b_l \\ \\
    0 & otherwise
    \end{array}\right.
    \label{eq:fs}
\end{equation}
where $a_l$, $b_l$, and $c_l$ are parameters to define a triangular membership function as shown in Figure \ref{fig:triangular}.

\begin{figure}[h]
    \centering
    \includegraphics[width=3.5in]{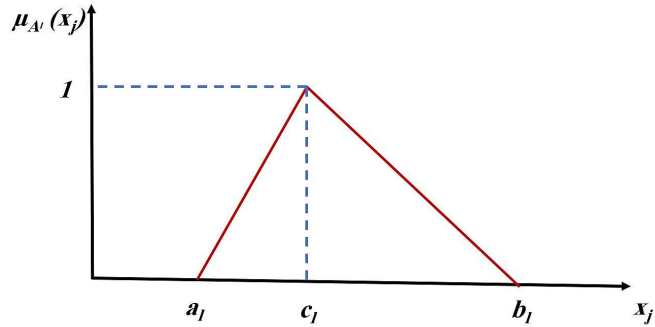}
    \caption{Triangular membership function.}
    \label{fig:triangular}
\end{figure}

\item{\textbf{Fuzzy Rules Layer}}: The neurons of this layer apply a \textit{T-Norm} operator on outputs of the previous layer to calculate rules' firing strength value. Indeed, each neuron of this layer represents a fuzzy rule and aggregates the membership values of input variables to fuzzy sets belongs to the antecedent part of each fuzzy rule like equation \ref{eq:rule}. In the proposed method the \textit{minimum} function is chosen as the \textit{T-Norm} operator. Therefore, the output of $i^{th}$ neuron as the firing strength of $i^{th}$ fuzzy rule is calculated as follows:

\begin{equation}
    f_i = min_j(\mu_{A_{j}^i}(x_j))
    \label{eq:FuzzR}
\end{equation}
where $\mu_{A_{j}^i}(x_j)$ indicates the membership value of $j^{th}$ input variable to the fuzzy set involved in $i^{th}$ rule related to the $j^{th}$ dimension ($A_{j}^i$), computed as an output value of the previous layer's neurons.

\item{\textbf{Normalization Layer}}: The neurons of this layer calculate the rules' normalized firing strength as follows:
\begin{equation}
    \phi_i = \frac{f_i}{\sum_{l=1}^{R}f_l}
    \label{eq:NFuzzR}
\end{equation}
where $R$ is the number of fuzzy rules.

\item{\textbf{Output Layer}}: Finally the output layer's neuron calculates the deffuzified final degree which would be utilized in scheduling algorithm (see section \ref{sec:alg}) as follows:
    \begin{equation}
    y = \sum_{l=1}^{R}\phi_l y_l
    \label{eq:output}
\end{equation}
where $y_l$ is the proposed output of the $l^{th}$ rule (see equation \ref{eq:rule}).
\end{itemize}

The details of the function of utilized FNN is summarized in Algorithm~\ref{alg:FNN}.

\begin{algorithm}[!t]
\scriptsize
  \textbf{Inputs:} \\
  $\quad$ \text{objective values for a processing core (c) and operational frequency (f):}\\
  $\quad\,$ \text{utilization (u), GSFR ($\Lambda$), Power (P), Temperature ($\theta$)} \\
  $\quad$ \text{Number of fuzzy sets ($N_{FS}$), Number of rules (R)} \\
  \textbf{Output:}\\
   $\quad\,$ \text{Degree of the determined c and f as their assignment appropriateness (y)}

\caption{The proposed FNN for task scheduling}
\label{alg:FNN}
    \vspace {2mm}
    - \textbf{Input Layer}:\\
    $\quad$ $x_1 \gets u$, $x_2 \gets P$, $x_3 \gets \theta$, $x_4 \gets \Lambda$ \\

    - \textbf{Fuzzification Layer}:\\
    \For{j=1:4}{
    \For{l=1:$N_{FS}$}{
    compute $\mu_{A_l}(x_j)$ based on eq. \eqref{eq:fs}
    }
    }

    - \textbf{Fuzzy Rules Layer}:\\
    \For{i=1:R}{
    compute fuzzy strength of the $i^{th}$ rule ($f_i$) based on eq. \eqref{eq:FuzzR}
    }

    - \textbf{Normalization Layer}:\\
    \For{i=1:R}{
    compute the normalized fuzzy strength of the $i^{th}$ rule ($\phi_i$) based on eq. \eqref{eq:NFuzzR}
    }

    - \textbf{Output Layer}:\\
    compute the final degree (y) based on eq. \eqref{eq:output}

\end{algorithm}

\subsection{Learning method}
\label{sec:learning}
To utilize the proposed fuzzy neural network, two problems should be first addressed: 1- structure identification, and 2- parameter estimation \cite{SOFMLS,Ebadzadeh15,Ebadzadeh2017}. The structure identification aims at determining the number of fuzzy sets and fuzzy rules for solving the problem. However, there are  antecedent and consequent parameters for each fuzzy rule that should be determined properly.

\begin{figure}[h]
    \centering
    \includegraphics[width=4.5in]{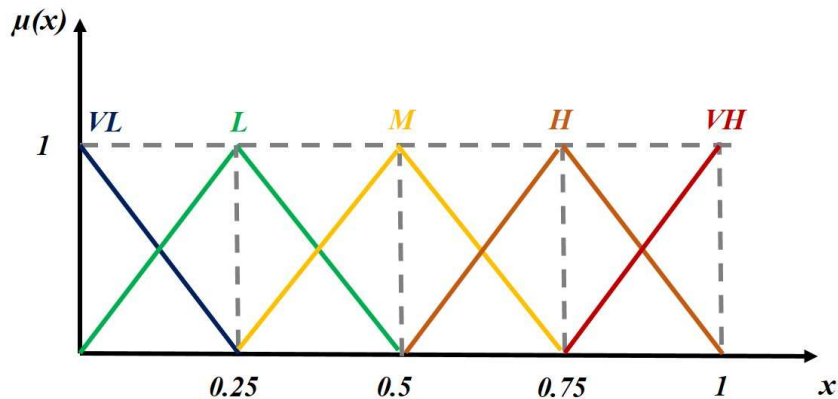}
    \caption{Showing the uniformly partitioning of input variables. The domain of each input variable x is partitioned to 5 fuzzy sets: Very Low level (VL), Low level (L), Medium level (M), High level (H), and Very High level (VH). The input variable is normalized to range [0,1] by dividing its value to its maximum value.}
    \label{fig:fs}
\end{figure}

There are some usual approaches to solve the \textit{structure identification} problem including using different clustering methods \cite{Malek11,Ebadzadeh15,Ebadzadeh2017,SALIMIBADR2022108258}, online clustering methods \cite{DFNN,GDFNN,DENFIS,SOFMLS,Salimi-Badr2017,AsadiEydivand2015,SalimiBadr2022}, and uniformly partitioning \cite{ANFIS,Ekhtiyari2019}.
The proposed structure is a control unit that should cover the whole input space properly. It is necessary to have one rule for each combination of different values of different variables. Therefore, it is not expected to determine fuzzy sets and fuzzy rules based on data-driven clustering approaches. Moreover, in the scheduling problem, the range of each input variable is predetermined. Therefore, we first normalize each input variable to the range of [0,1]. Next we partition the range of each input variable uniformly. Figure \ref{fig:fs} shows the proposed uniformly partitioning of input variable $x$.

Uniformly partitioning of the input variables provides the values of the premise parts' parameters ($a_i$, $b_i$, and $c_i$ (i=1,2,...,R) in equation \eqref{eq:fs}). Therefore, the only parameter that its value should set properly is the consequent part's parameter of each fuzzy rule ($y^i$ in equations \eqref{eq:rule} and \eqref{eq:output}). The value of this variable determines the rank of various voltage/frequency levels of each core to assign the current task in the scheduling and mapping procedure (see section \ref{sec:alg}).

An usual approach to learn consequent parts' parameters of a fuzzy neural network is to minimize the error between the desired and actual output values using local search algorithms including \textit{Gradient Descent} (GD), \textit{Levenberg-Marquardt} (LM), or Linear Least Square Error (LLS) methods \cite{Ebadzadeh09,Ebadzadeh15,Ebadzadeh2017,SALIMIBADR2022108258,SALIMIBADR2022139}. Here, the desired output values (the "labels" in the supervised learning) are not available. Therefore, these local search methods are not applicable.

Instead of using the error between the desired and actual output values, here we can evaluate the total performance of the fuzzy neural network during the scheduling and mapping process. Indeed, we propose to derive the consequent parts' parameters based on maximizing the performance of the fuzzy neural network in task scheduling and mapping process. Therefore, we should consider some criteria to measure the performance of the model. Afterward, an optimization approach should determine the consequent parts' parameters based on maximizing the performance criteria.

There are several criteria for measuring the performance of a scheduling and mapping algorithm. Some of critical objectives are \textit{utilization}, \textit{power}, \textit{core temperature}, \textit{global system failure rate ($GSFR$))}, (described in section \ref{sec:objectives}) \cite{HYSTERI,MEJ,ERPOT}. Therefore, to learn proper values of the consequent parts' parameters a multi-objective optimization problem should be addressed to consider all mentioned important criteria.

An efficient approach to solve a multi-objective optimization problem is the  \textit{NSGA-II} algorithm \cite{nsga}. NSGA-II is an evolutionary algorithm that considers several objectives in the sorting and selecting the next generation individuals. This algorithm provides non-dominant solutions by forming a Pareto front. NSGA-II has been successfully applied to the scheduling problem \cite{MEJ}.

To use the NSGA-II the fitness of each individual is defined based on minimizing the cost presented as follows:
\begin{equation}
   cost = [(U(C)),\theta(C),P(C)), \Lambda(C))]
    \label{eq:cost}
\end{equation}
where, $U(C)$, $\theta(C)$, $P(C)$, and $\Lambda(C)$ are utilization, temperature, power, and global system failure rate ($GSFR$) of the core $C$.

\begin{figure}[h]
    \centering
    \includegraphics[width=3in]{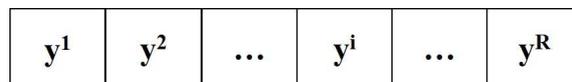}
    \caption{The structure of each chromosome representing a.solution (a Fuzzy Neural Network). The consequent parts' parameters ($y^{i}$, i=1,2,...,R) are considered as the genes of each chromosome.}
    \label{fig:chromosome}
\end{figure}

To learn the consequent parts parameters, each individual is considered as a fuzzy neural network. Since the premise parts' parameters are determined based on the uniformly partitioning, to encode a fuzzy neural network only the consequent parts' parameters are considered as the genes of each chromosome as shown in Figure \ref{fig:chromosome}.

The NSGA-II final result would be a Pareto front that consists of non-dominant optimal solutions. Indeed, the final result of this evolutionary engine is a set of proper solutions. To extract one solution from the extracted Pareto front we choose the point with minimum distance from the other solutions. In this way, the middle point of the Pareto front would be selected that considers all criteria with almost equal weights.

To learn the consequent parts' parameters, the NSGA-II algorithm is applied to a training dataset including different application graphs with various size and attributes. For each benchmark a Pareto front is extracted and a solution is chosen from all extracted valid solutions.  Next, the final solution is calculated by averaging the chosen solutions of different training benchmark problems. The learning algorithm is summarized in Algorithm \ref{alg:learning}.

It is worthy to mention that all rules of a solution don't fire during solving a problem. Indeed it is possible that during scheduling an application, the premise parts of some rules (mainly the extreme rules) are not be satisfied. Consequently the fitness of all rules can not show the qualification of these inactivated rules. Therefore the inactivated rules are not considered in extracting the final solution.

\begin{algorithm}[!t]
\scriptsize
  \textbf{Inputs:} \\
  $\quad$ \text{Number of Rules ($R$), Premise Parts Parameters ($a_i$,$b_i$, and $c_i$)}\\
  $\quad$ \text{Iteration Number, Population size ($N_{pop}$)} \\
  $\quad$ \text{Crossover Probability ($P_c$), Mutation Probability ($P_m$)} \\
  $\quad$ \text{Training set $T$ including DAGs of different applications} \\
  $\quad$ \text{Architecture graph of heterogeneous MPSoC } \\
  \textbf{Output:}\\
   $\quad\,$ \text{$y_{out}$: Final Consequent Parts' Parameters of FNN ($y^i$)}\\
\caption{Learning Algorithm}\label{alg:learning}
\vspace{2mm}
    $y \gets zeros(R)$\;
    \For{$t \in T$}{
            \text{Extract Pareto Front applying NSGA-II considering cost functions in eq.~\eqref{eq:cost}}:

            \text{Choose the middle point of Pareto Front ($y_{mid}$)}:

                \For{$p_i , p_j \in$ Pareto Front}{
                    Distance [i,j] $\gets \parallel p_i-p_j\parallel^2_2$
                }

                $y_{mid} \gets y[argmin_{i} Distance]$ \\
                $y_{out} \gets y_{out} + y_{mid}$
                }
                $y_{out} \gets y_{out}/length(T)$
\end{algorithm}

\section{Experimental Results}
\label{sec:results}

In this section the effectiveness of our proposed task scheduling approach in joint optimizing the main design challenges of heterogeneous MPSoC is demonstrated by several experiments.
These experiments are classified to five categories based on their various targets and are performed on a simulated heterogeneous MPSoC platform executing several real-life and synthetic benchmarks.

\subsection{Simulation Setup}
\label{sec:simset}

Our proposed evolutionary-based neurofuzzy task scheduling approach is implemented on a simulated heterogeneous MPSoC platform consists of four processing cores each equipped by three levels of operational voltage and frequency levels. The considered operational voltage and frequency sets are presented in~\ref{table:param}.
These heterogeneous processing core are different in processing capability and appropriateness on executing tasks based on their properties. This simulation platform is implemented by MATLAB R2018a and the characteristics of considered processing cores are adopted from real instances such as ARM cortex processors~\cite{cortex_2011}. Moreover, in the experiments several real-life and synthetic benchmarks in form of application graphs are considered. The real-life benchmarks are selected from E3S benchmark suits and the synthetic ones are randomly generated by TGFF tool~\cite{E3S,tgff}. The sizes of these graphs are set from 8 nodes to 100 nodes to cover various application types and evaluate the effectiveness of the proposed approach in different scenarios.

To simulated our proposed task scheduling approach, we have implemented an evolutionary-based optimization engine based on NSGA-II to determine the appropriate fuzzy rules. To this aim, our defined multi-objective optimization problem is solved by this engine. The considered parameters of NSGA-II algorithm is presented in table~\ref{table:param}. In this context, to compute the lifetime reliability, power consumption and temperature as the main objectives, their static models that are presented in section~\ref{sec:pre} are utilized. The constant parameters of these models are also summarized in table~\ref{table:param} and are mainly adapted from~\cite{thiele_2013,JEP122H}. 
The simulation platform is built on an Intel quad-core i7 CPU with
64GB RAM.

\begin{table}[]
    \centering
    \scriptsize
    \begin{tabular}{|c|c|}
        \hline
        \textbf{Parameter} &  \textbf{Values}\\
        \hline
        \text{Lifetime Reliability} & \makecell{$ E_a (EM,SM) = 10444.07, K = 8.61e-5, J =150, n_{EM} = 1.1, n_{SM} = 2.5,$ \\ $n_{NBTI} = 5, E_a (NBTI) = 4651.16, q_{TC} = 2.35,$ \\ $ \{a,b,x,y,z\}_{tddb} = \{78,-0.08,0.76,-66.8,-8.37e-4\} $} \\
        \hline
        \text{Power Consumption} &  \makecell{$C_{eff} = 10^{-8} J/V^2, F = \{300,600,900\} MHz, V = \{1.06,1.1,12\}$ \\  $\alpha = 0.1 W/K, \beta = -11 W/K V$} \\
        \hline
         \text{Temperature} & $C = 0.03 J/K, G = 0.3 W/K, T_{amb} = 293 K, G_{neighbor} = 0.1 W/K$ \\
         \hline
         \text{NSGA-II} & $N_{pop}=200, P_c = 40\%. P_m = 70\%, iteration number = 500$ \\
         \hline
    \end{tabular}
    \caption{Parameter values of the objective models and NSGA-II engine}
    \label{table:param}
\end{table}

\subsection{Simulation Results}
To evaluate the efficiency of our proposed evolutionary-based neurofuzzy task scheduling approach, in solving the target problem and comparing it to related studies, three classes of experiments are considered and presented in this section. First, the joint-optimization ability of the method is investigated by presenting the surface of the extracted Pareto front. Next, the interpretability of the model is studied by presenting the extracted fuzzy rules of the selected solution from the Pareto front. Afterward, the performance of the method is compared with some previous approaches in offline and online settings.

\begin{enumerate}[(I)]

\item \textbf{Joint Optimization Ability:} We have trained the model based on NSGA-II on a training dataset consisting of several synthetic and real-life application graphs with various sizes. It includes ten application graphs with sizes in range of 6 to 85 nodes. In order to show the joint optimization capability of the proposed method, the final Pareto front related to a graph with 40 nodes is shown in Figures \ref{fig:pareto1} and \ref{fig:pareto2}.
We have chosen this graph because its size is middle (not too large and not too small). Since the optimization problem considers four criteria at the same time, to show the four dimensional Pareto front in a figure, three axes are set as power, temperature, and GSFR and different values of the fourth dimension, the execution time, are represented by different colors of solution points.

\begin{figure}[h]
    \centering
    \includegraphics[width = 4 in]{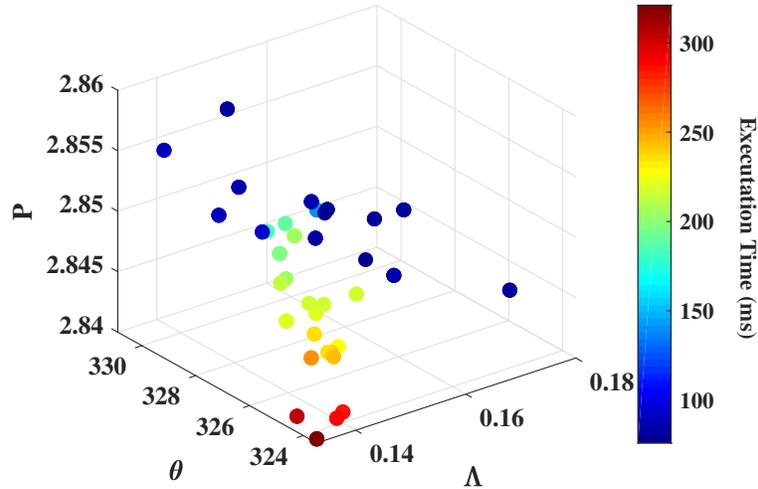}
    \caption{The extracted four-dimensional Pareto front. Three dimensions including temperature ($\theta$), power ($P$), and GSFR ($\Lambda$) are shown on axes and the fourth dimension, the execution time, is encoded in the colors.}
    \label{fig:pareto1}
\end{figure}

As this figure shows, there exists trade-offs among these design challenges and having low-temperature, power and failure rate solutions cost a high execution time (red points of Figure~\ref{fig:pareto1}). Reversely, blue points represents the solutions that are accessible with lower execution time but higher costs in terms of temperature and power consumption. As this Pareto front shows, our proposed approach is capable of exploring the whole design space and the extracted rules have an appropriate coverage on it. This coverage is reflected in various colors of points from the whole range of execution time spectrum. Thus, based on the occurred scenario and the application specifics, one of these solution points could be selected and applied to the system that properly meets the existing trade-off among design challenges during the task scheduling process.

Moreover, to demonstrate the appropriateness and coverage of the extracted rules by our proposed learning approach, Figure~\ref{fig:pareto2} shows the projection of each extracted solution point in two dimensions. In this figure, six various projections are considered based on our considered objectives (execution time, temperature, power consumption and failure rate). Besides, the points are labeled to make them traceable in various viewpoints and compare their corresponding values.

\begin{figure}
\centering
\begin{tabular}{cc}
    \subfigure[][]{\includegraphics[width=2.5 in]{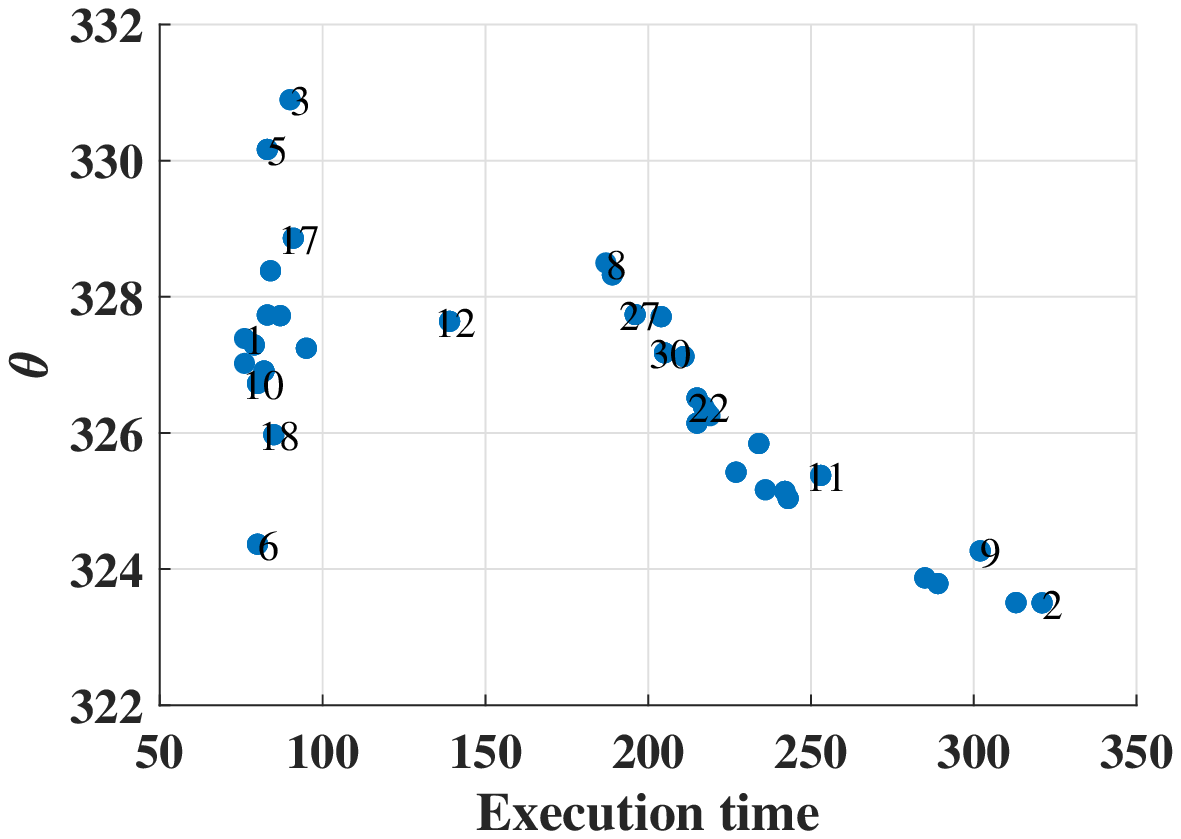}} &
    \subfigure[][]{\includegraphics[width = 2.5in]{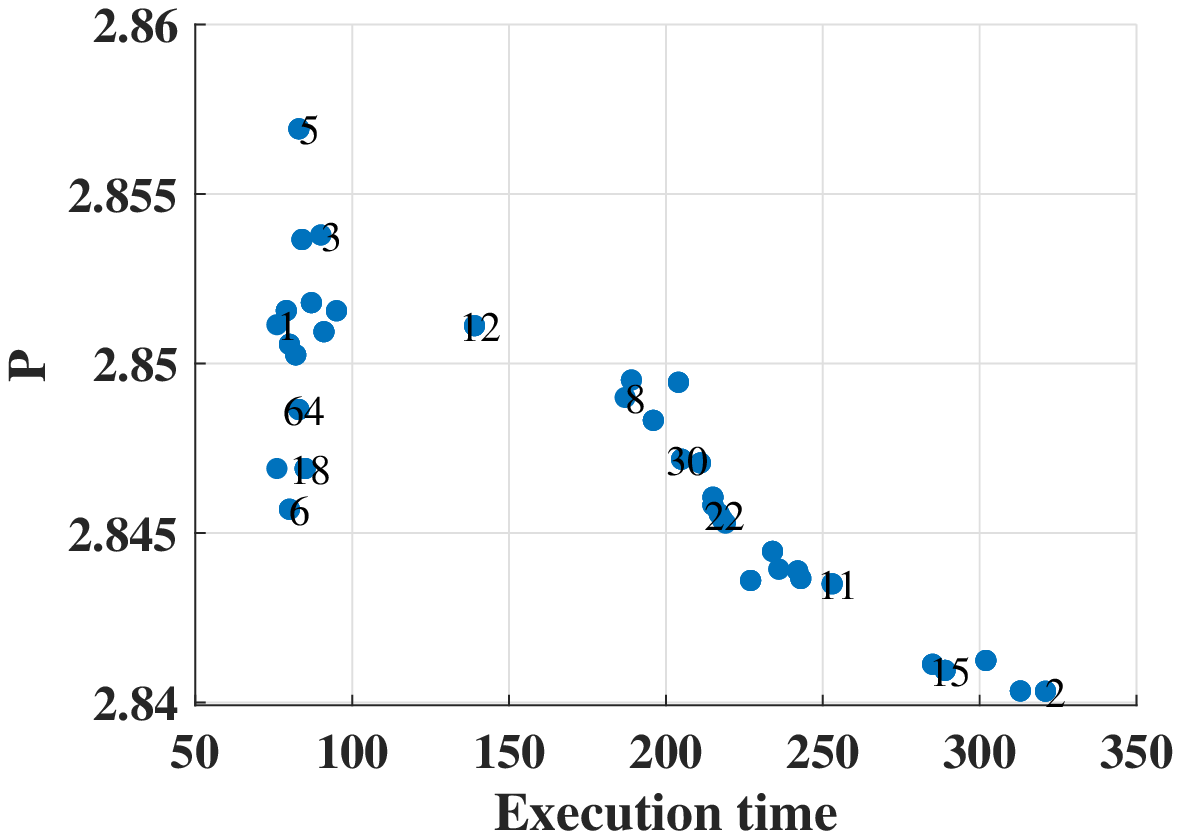}} \\
        \subfigure[][]{\includegraphics[width=2.5 in]{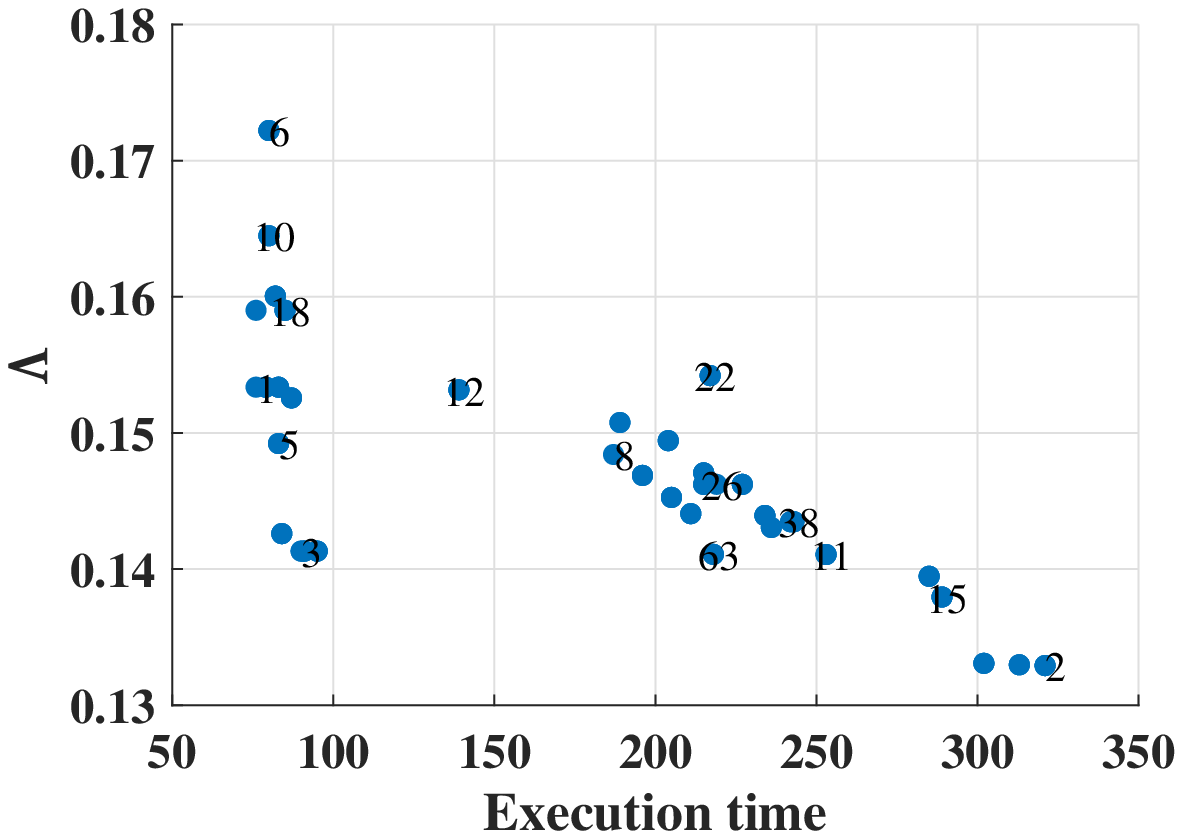}} &
    \subfigure[][]{\includegraphics[width = 2.5in]{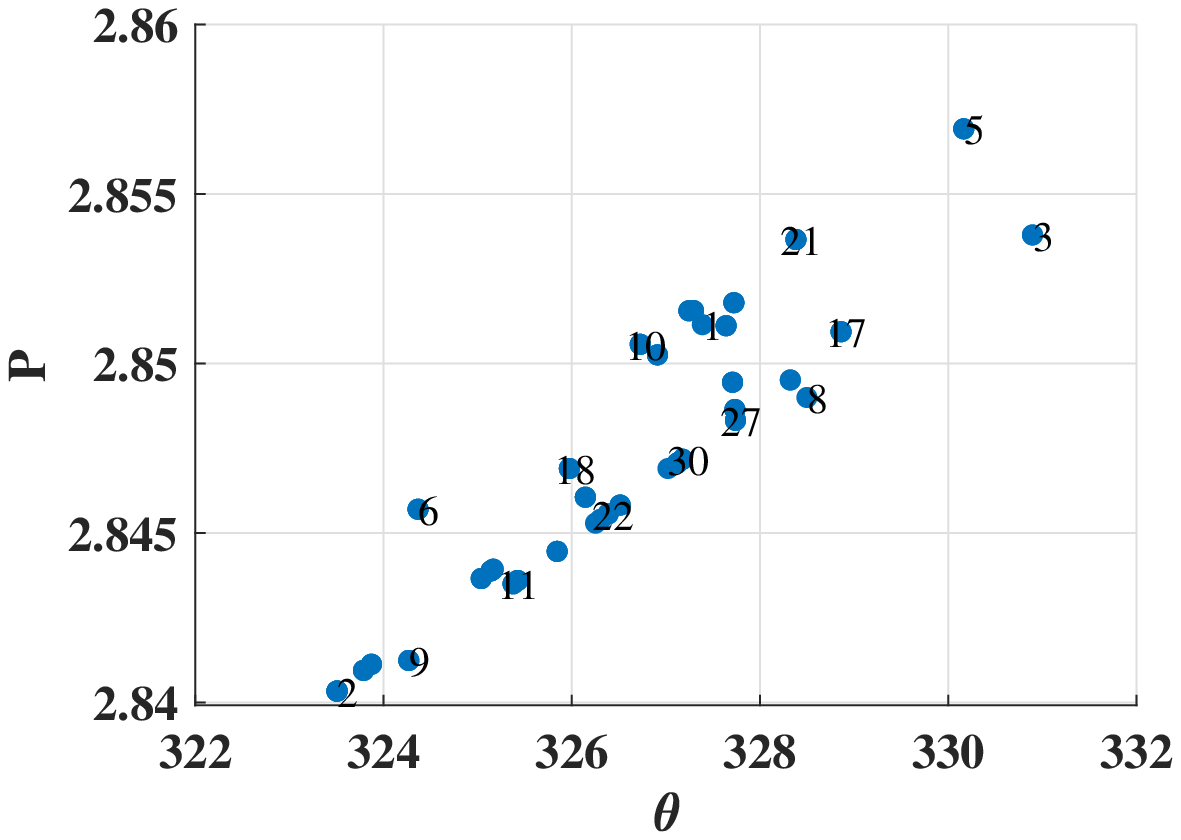}} \\
        \subfigure[][]{\includegraphics[width=2.5 in]{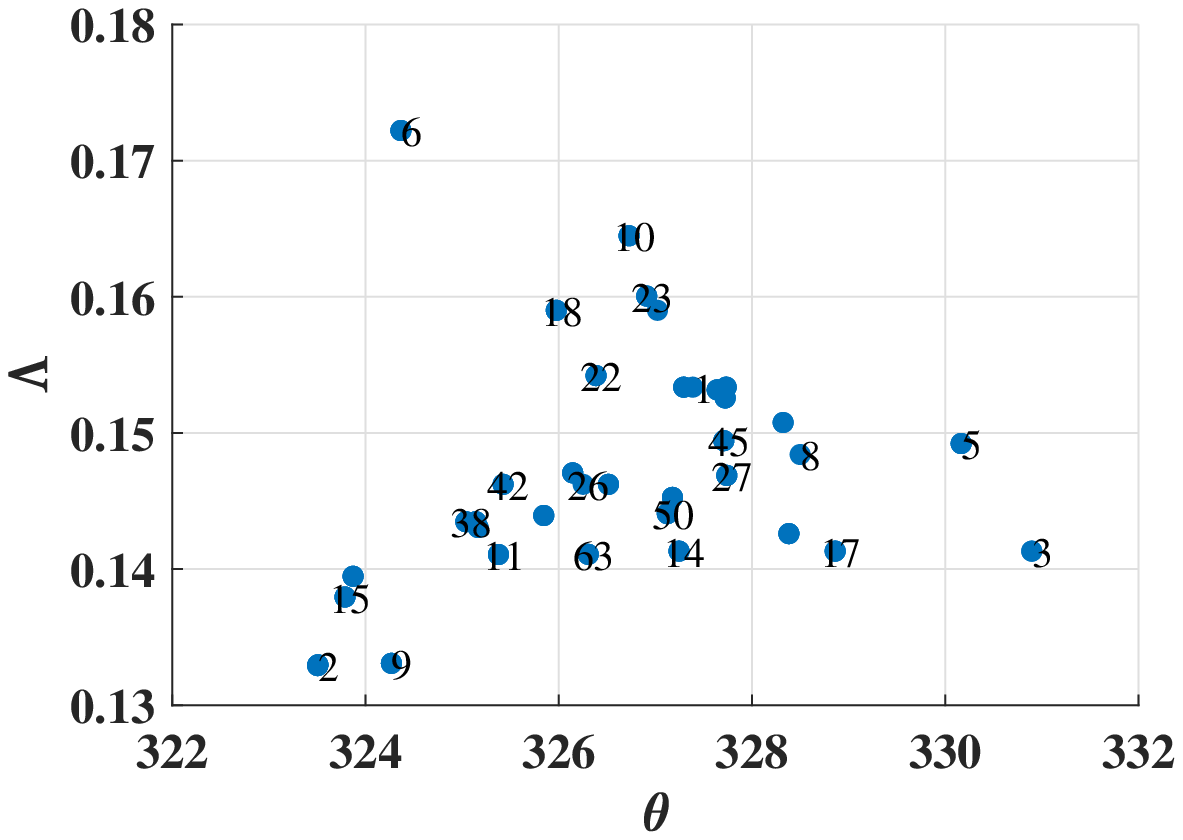}} &
    \subfigure[][]{\includegraphics[width = 2.5in]{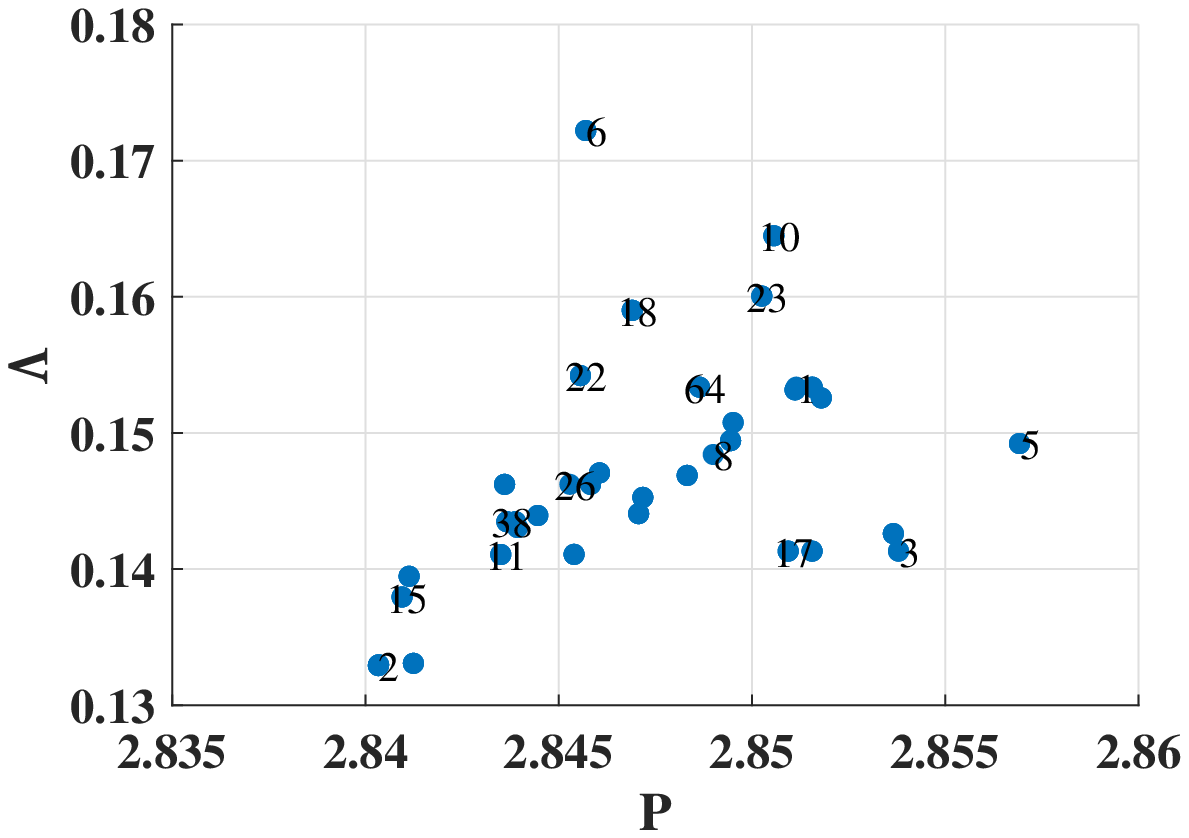}} \\
\end{tabular}
\caption{Projections of the four-dimensional Pareto front to the various two-dimensional planes. Some solutions are labeled with numbers to simplify the tracking of different solutions in different diagrams.}
\label{fig:pareto2}
\end{figure}

\item \textbf{Extracted Fuzzy Rules:} We have selected one solution from those extracted as the Pareto front based on the process explained in section \ref{sec:learning} and Algorithm \ref{alg:learning}. According to the uniform partitioning of four input variables to five fuzzy sets, we have 625 extracted rules. Consequently, it is not possible to show them all clearly. Therefore, we have selected the most frequently fired fuzzy rules during the scheduling process. To select these rules, the average of firing strength values of each fuzzy rule during scheduling of different application graphs of the training dataset is calculated. Next, the rules with mean firing strength value greater than 0.1 are selected and shown as a heat-map in Figure \ref{fig:rules}. In this figure, the color spectrum reveals the linguistic values of different variables as our our objectives (from blue to red indicates the lower values to higher ones).

\begin{figure}[h]
    \centering
    \includegraphics[width = 4 in]{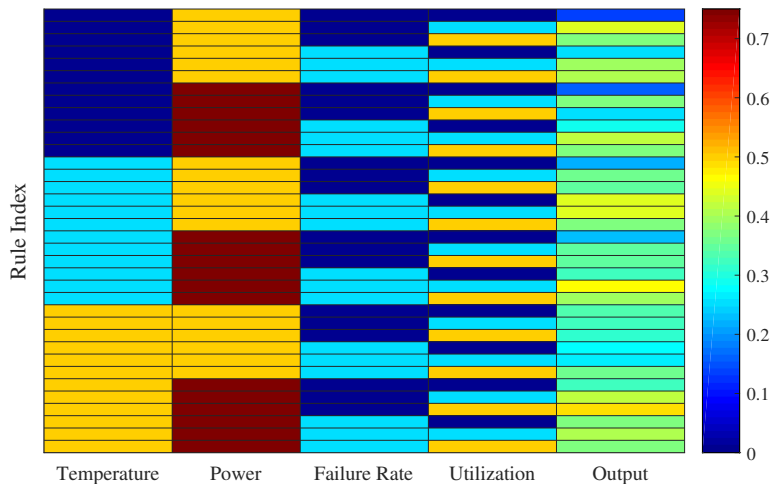}
    \caption{The most frequently fired fuzzy rules extracted from our proposed learning procedure. Columns are the input/output variables and the rows indicate the fired rules' indexes. The colors from blue to red indicates the lower values to the higher ones}
    \label{fig:rules}
\end{figure}

As this figure shows, satisfaction of the existing trade-off among the main objectives of our target scheduling problem. This trade-off is reflected in the color of output column (middle range of colormap). Where all the objectives are at an appropriate levels (blue range of the colormap), the value of the output is higher (end of middle part of colormap). This high value leads to increase the chance of selecting these cores during the task scheduling process due to their appropriateness in optimizing the main design challenges of the system.
Moreover, based on the various values of the input columns, that are distributed in the colors of this figure, it is seen that our learning process extracts the rules that explore the design space appropriately. Furthermore, the extreme values that contradict the mentioned trade-off are not fired frequently hence they are not included in this figure.

\item \textbf{Comparison with Previous Methods:}
To compare the efficiency of our proposed task scheduling approach to related studies, first four heuristic and meta-heuristic methods as the most related studies are considered. ERPOT and PowerPerf-PET~\cite{ERPOT,sheikh_2016sixteen} are considered as offline heuristic task scheduling approaches. In~\cite{ERPOT}, the joint optimization of the main design challenges of MPSoCs is considered during a static list scheduling-based Pareto-based optimization approach. Here, three out of four objectives are transformed to the constraints and for their various target values, the multi-objective optimization is solved in form of a single-objective problem. In~\cite{sheikh_2016sixteen}, the optimization objectives are summing up with an appropriate weight. Moreover, the mapping process is performed through a greedy approach to minimize the product of power consumption and execution time of each task. This approach does not consider lifetime reliability as an objective during the task scheduling process and the thermal effect of neighbors on an MPSoC is also discarded. To have a fair comparison, we have added these characteristics to PowerPerf-PET method during this experiment.
Moreover, two meta-heuristic-based task scheduling approaches are also selected for comparing to our proposed method. They are based on NSGA-II and SPEA-II as the most effective evolutionary-based multi-objective optimization approaches~\cite{MEJ,spea_2}. These approaches explores the design space and generate a Pareto front for each application graph to jointly optimize the main design challenges of the target heterogeneous MPSoC. Table~\ref{table:comp} shows the results of comparing our proposed approach to the described selected methods. This comparison is performed for ten synthetic and real-life application graphs with various sizes of 6 to 85 tasks.Lifetime reliability, temperature, power consumption and execution time values are studied separately in this table. Moreover, the average values of all distinct solutions of the Pareto front in various aspects are reported in this table.

\begin{table}[t]
    \centering
    \scriptsize
    \resizebox{\columnwidth}{!}{
    \begin{tabular}{|c|cccc|cccc|cccc|cccc|}
        \hline
        & \multicolumn{4}{c}{ERPOT} \vline &\multicolumn{4}{c}{PowerPerf-PET} \vline & \multicolumn{4}{c}{NSGA-II} \vline & \multicolumn{4}{c}{Proposed method} \vline \\
    \hline
    \hline
        & $\theta$ & $P$ & $\Lambda$ & $E$ & $\theta$&$P$&$\Lambda$&$E$ & $\theta$&$P$&$\Lambda$&$E$ & $\theta$&$P$&$\Lambda$&$E$ \\
        \hline
        Auto-industry & 313.14 & 2.05 & 1.95e-4 & 86 & 317.13 & 2.43 & 2.3e-4 & 135 & 312.6 & 2.3 & 1.06e-4 & 84 & 316.02 & 2.29 & 6.14e-5 & 82 \\
        \hline
        Random-85 & 373.58 & 3.22 & 1.01e-3 & 409 & 377.3 & 3.84 & 2.66e-3 & 655 & 368.1 & 3.19 & 2.5e-4 & 521 &  346.35 & 2.86 & 1.57e-4& 413\\
        \hline
        Gaussian & 312.85 & 2.27 & 2.4e-5 & 76 & 315.4 & 2.31 & 3e-5 & 130 & 311.12 & 2.9 & 2.13e-5 & 76 & 312.72 & 2.42 & 1.7e-5 & 73\\
        \hline
        Random-40 & 339.07 & 2.91 & 2.3e-4 & 98 & 332.3 & 3.00 & 2.13 e-4 & 117 & 337.5 & 3.01 & 1.59e-4 & 88 & 324.58 & 2.79 & 1.12e-5 & 99 \\
        \hline
        Random-50 & 338.72 & 3.16 & 2.64e-4 & 182 & 343.8 & 3.41 & 6.8e-4 & 264 & 332.5 & 2.87 & 1.89e-4 & 155 & 330.04 & 2.87 & 1.27e-5 & 182\\
        \hline
        Office & 313.18 & 2.84 & 2.75e-5 & 62 & 324.3 & 3.04 & 2.1e-4 & 95 & 314.3 & 2.54 & 2.53e-5 & 56 & 309.65 & 2.77 & 1.02e-5 & 43 \\
        \hline
        Network & 313.88 & 2.84 & 5.18e-5 & 62 & 324.0 & 3.09 & 2.08e-4 & 97 & 315.8 & 2.71 & 2.97e-5 & 66 & 308.95 & 2.76 & 1.02e-5 & 58\\
        \hline
        Random-64 & 370.05 & 3.72 & 1.004e-3 & 294 & 382.3 & 4.03 & 1.95e-3 & 433 & 366.9 & 3.45 & 6.9e-4 & 289 & 369.19 & 2.87 & 1.9e-4 & 301\\
        \hline
        Random-49 & 342.07 & 3.32 & 3.39e-4 & 255 & 348.5 & 3.79 & 7e-4 & 355 & 356.9 & 3.12 & 2.97e-4 & 266 & 349.96 & 2.79 & 1.5e-4 & 199\\
        \hline
        Random-28 & 328.82 & 2.51 & 1.76e-4 & 191 & 337.96 & 3.3 & 4e-4 & 251 & 326.2 & 2.43 & 1.69e-4 & 196 & 329.12 & 2.63 & 2.51e-5 & 172\\
        \hline
    \end{tabular}}
    \caption{Comparison of the proposed method to ERPOT, PowerPerf-PET, NSGA-II-based and SPEA-II-based approaches as the most  related studies in terms of temperature, power consumption, failure rate and execution time}
    \label{table:comp}
\end{table}

The second comparison is performed to a fuzzy-based offline scheduling approach, that utilized a static fuzzy inference system to jointly optimize utilization, power consumption and temperature of heterogeneous MPSoCs~\cite{Ekhtiyari2019}. In this method, the degree of processors during task scheduling and mapping are determined based on the FIS and and its rules that are statically perform some kind of biased voting toward the lower values among the main optimization objectives. Figure~\ref{fig:compbeit} shows the comparison results of our proposed FNN-based task scheduling approach to~\cite{Ekhtiyari2019} in terms of the target main design challenges of MPSoCs. Here, the same as previous experiment, ten synthetic and real-life application graphs are considered and the distribution of normalized values for each considered objective for all application graphs are compared.

\begin{figure}[t]
    \centering
    \includegraphics[width = 4in]{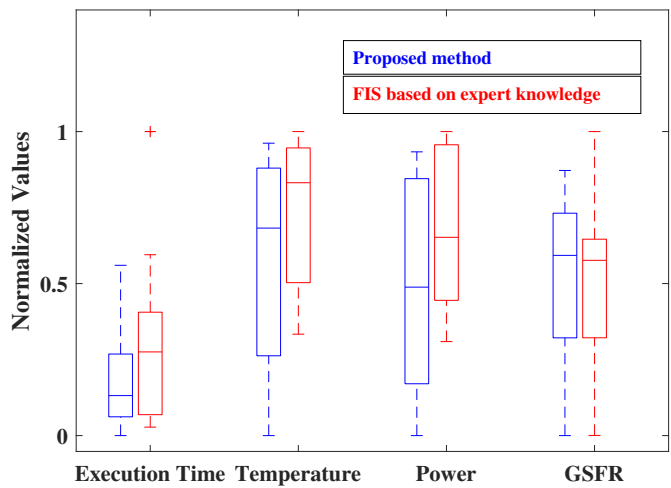}
    \caption{Statistical comparison of the proposed method with an FIS based on expert knowledge \cite{Ekhtiyari2019} according to different criteria.}
    \label{fig:compbeit}
\end{figure}

The detailed comparison of our proposed task scheduling approach to the FIS-based scheduling of~\cite{Ekhtiyari2019} is presented in Figure~\ref{fig:compbar}. In this figure, the optimization capability of these approaches is compared for each considered objective. As this figure shows, our proposed FNN-based task scheduling approach outperforms the FIS-based scheduling of~\cite{Ekhtiyari2019} in terms of all parameters. This improvement is achieved due to the employed learning-based rule set of our approach that is generated based on joint optimization of the design challenges rather than the static rule set of~\cite{Ekhtiyari2019}. It should be mention that, this experiment is performed on ten various synthetic and real-life application graphs with various sizes that are selected from~\cite{tgff,E3S}.

\begin{figure}[h]
\centering
\begin{tabular}{cc}
    \subfigure[][]{\includegraphics[width=2.7 in]{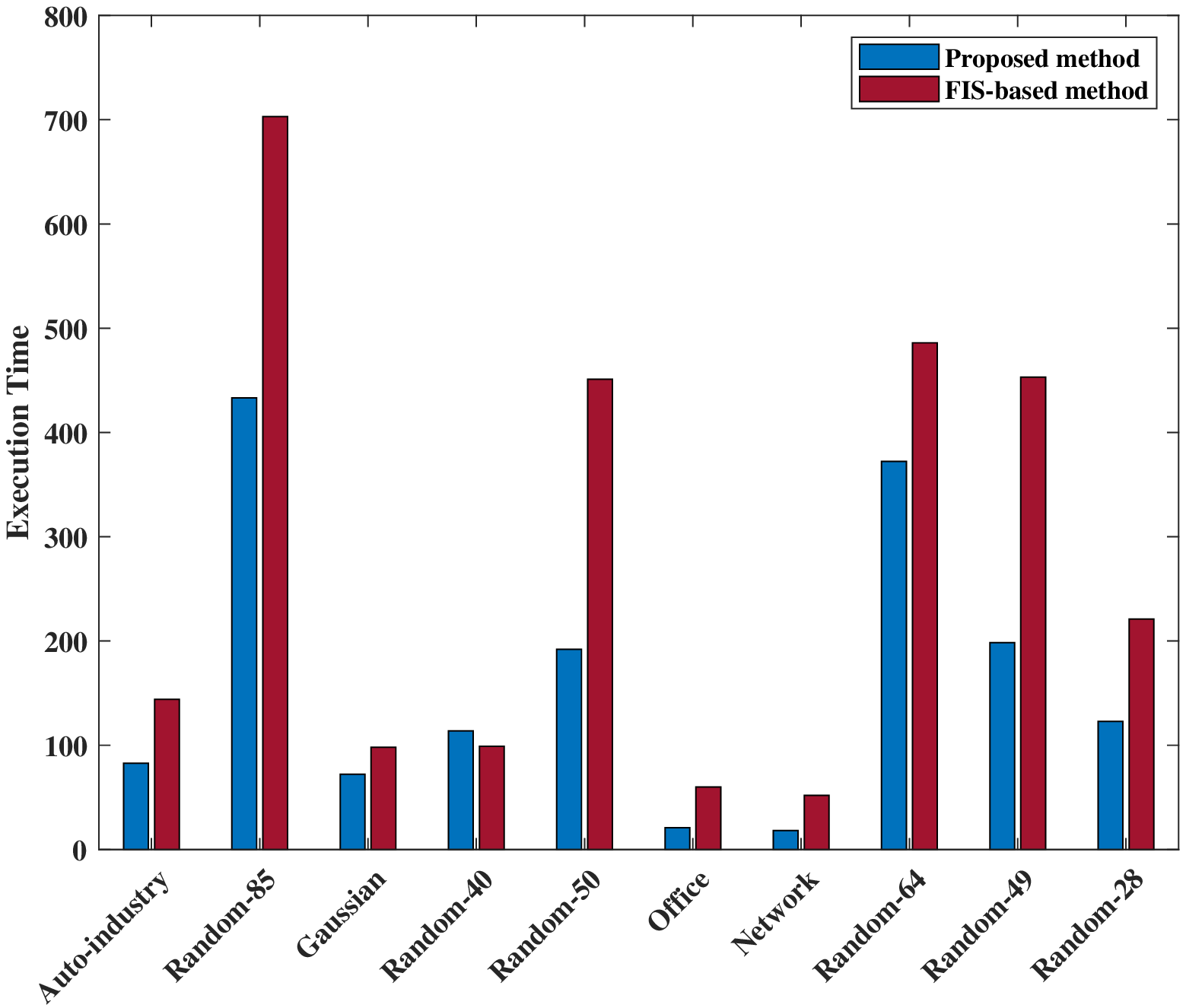}} &
    \subfigure[][]{\includegraphics[width = 2.7in]{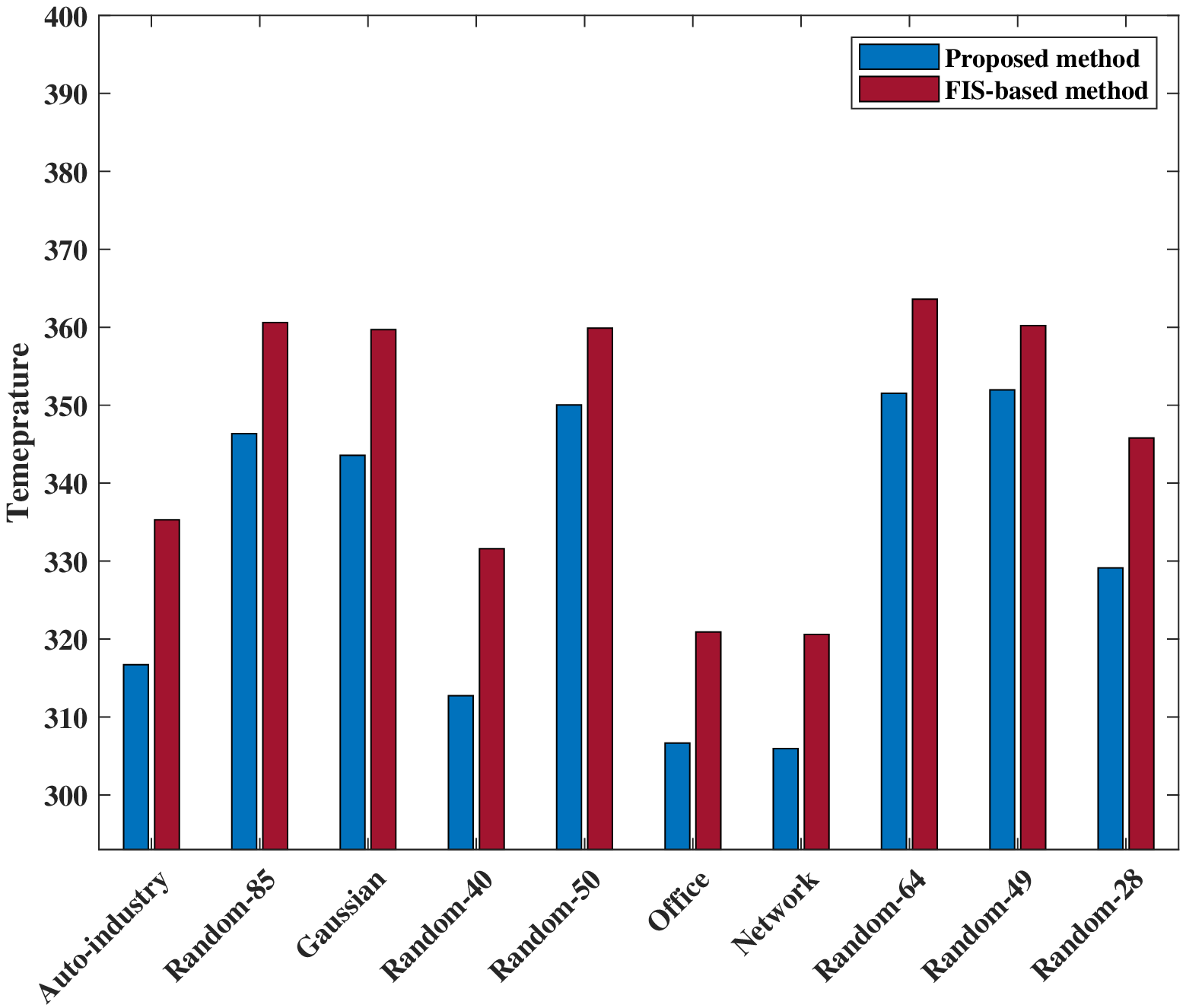}} \\
        \subfigure[][]{\includegraphics[width=2.7 in]{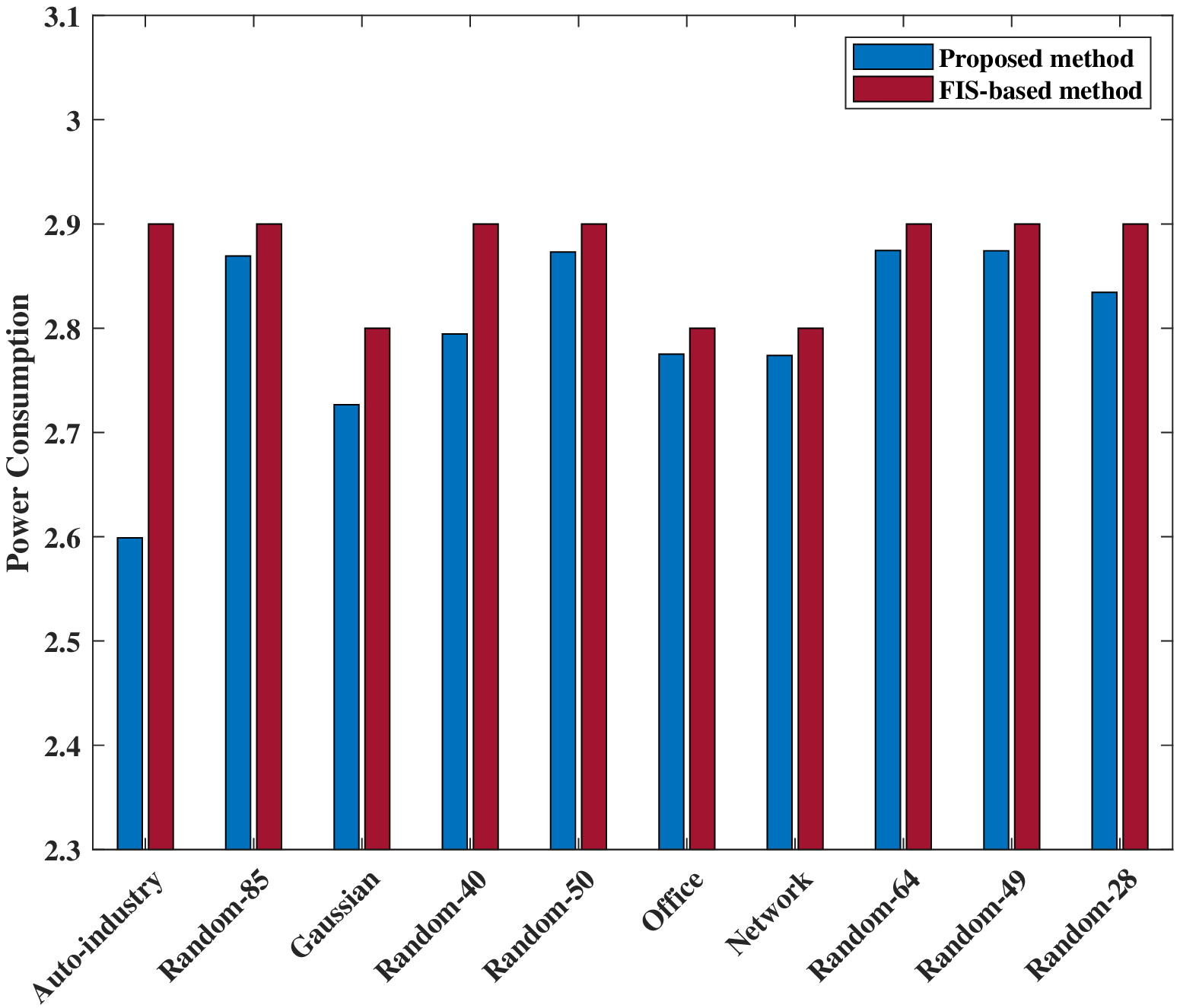}} &
    \subfigure[][]{\includegraphics[width = 2.7in]{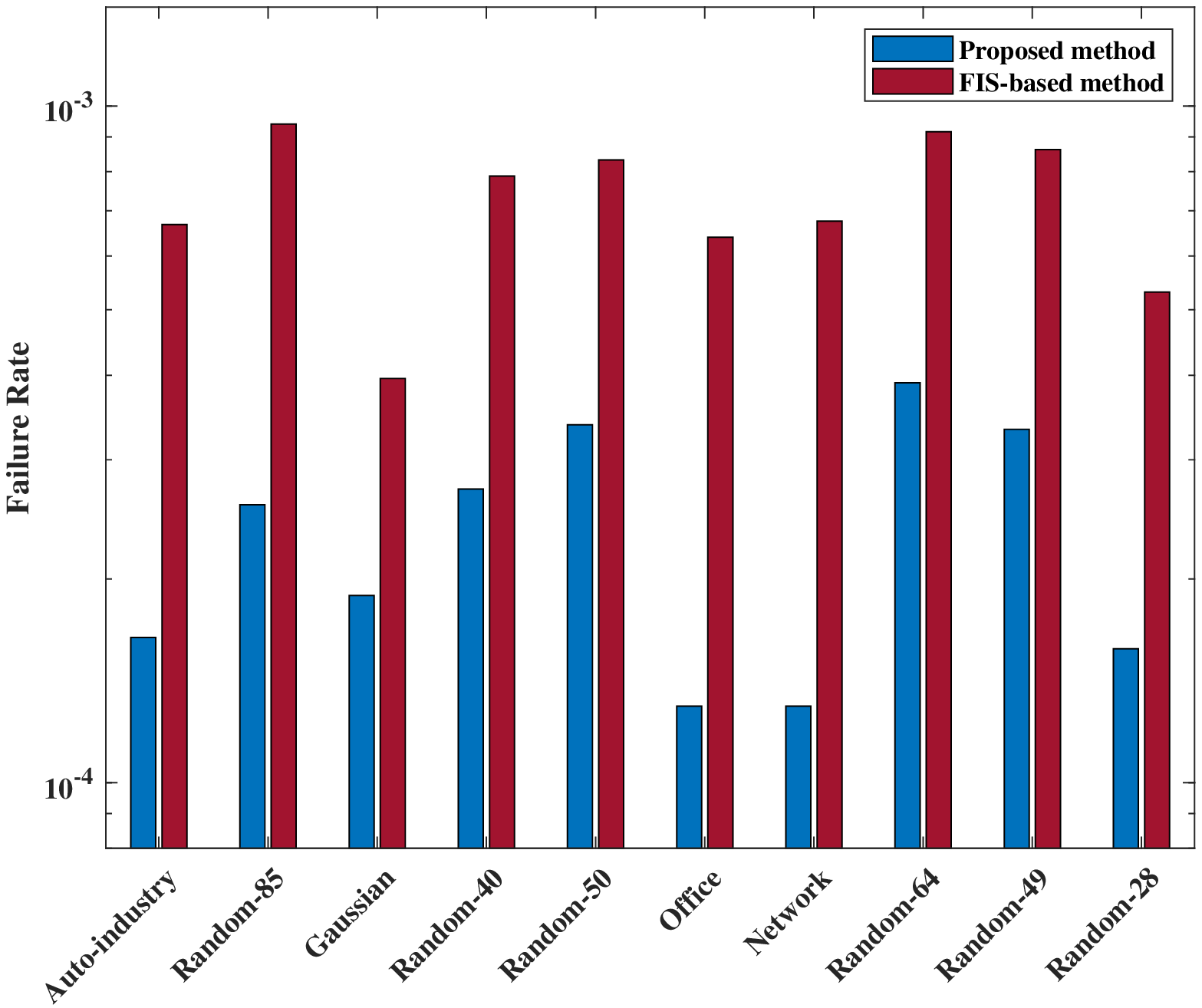}} \\
\end{tabular}
\caption{Comparison of our proposed FNN-based task scheduling approach to the FIS-based method of ~\cite{Ekhtiyari2019} in terms of execution time, temperature, power consumption and failure rate}
\label{fig:compbar}
\end{figure}
\end{enumerate}

\section{Conclusions}
\label{sec:concusion}
In this paper, an evolutionary-based neuro-fuzzy task scheduling approach is proposed. The method is learned using a multi-objective evolutionary algorithm (NSGA-II) to jointly optimize the main design challenges of heterogeneous MPSoCs, including temperature, power consumption, failure rate, and execution time.

Considering the noise in sensor measurements and the difference between computational models and reality, the task scheduling process encounters uncertainty. Therefore, it is expected to overcome this uncertainty by utilizing fuzzy neural networks. Moreover, another advantage of using a fuzzy neural network is to extract interpretable fuzzy rules that are understandable for an expert.

The final model is a rule-based online scheduler that receives the state of each core (temperature, power consumption, failure rate, and utilization of each core) as the input and calculates an appropriateness degree for assigning the current task to that core. Next, the current ready task is assigned to the core with the highest value of appropriateness degree.
Uniform partitioning of the input space is used to adjust antecedent parts' parameters of fuzzy rules. Moreover, NSGA-II is used to learn consequent parts' parameters of fuzzy rules based on considering the main
design challenges of MPSoCs.

To demonstrate the efficiency of the proposed method, several experiments on synthetic and real-life application graphs are performed. In this context, the final solution set is extracted in form of a Pareto front and its capability in meeting the existing trade-offs among various design challenges is studied. Moreover, the most frequently fired extracted fuzzy rules are presented to show the interpretability of the model. Finally, the performance of the proposed approach in optimizing the main design challenges is compared with some previous heuristic, meta-heuristic, and rule-based methods. According to these experimental results, the proposed method efficiently outperforms the previous models in all criteria.
As a future trend, expanding the proposed method using higher-type fuzzy logic including the interval type-2 to better consider the uncertainty is proposed.

 \bibliographystyle{plain}
 \bibliography{cas-refs}

\newcommand{\noop}[1]{}
\begin{thebibliography}{10}

\bibitem{E3S}
Embedded system synthesis benchmark suite ({E3S}).
\newblock \--url{http://ziyang.eecs.umich.edu/~dickrp/e3s/}.
\newblock Accessed: 2022-02-09.

\bibitem{scheduling_book_2006}
{\em Multicriteria Scheduling}.
\newblock Springer-Verlag, 2006.

\bibitem{abdallah2017advanced}
Abderazek~Ben Abdallah.
\newblock {\em Advanced Multicore Systems-On-Chip}.
\newblock Springer, 2017.

\bibitem{soft_2020_whale}
Mohamed Abdel-Basset, Doaa El-Shahat, Kalyanmoy Deb, and Mohamed Abouhawwash.
\newblock Energy-aware whale optimization algorithm for real-time task
  scheduling in multiprocessor systems.
\newblock {\em Applied Soft Computing}, 93:106349, 2020.

\bibitem{ERPOT}
athena Abdi, Alain Girault, and Hamid~R. Zarandi.
\newblock Erpot: A quad-criteria scheduling heuristic to optimize execution
  time, reliability, power consumption and temperature in multicores.
\newblock {\em IEEE Transactions on Parallel and Distributed Systems},
  30(10):2193--2210, 2019.

\bibitem{HYSTERI}
Athena Abdi and Hamid~R. Zarandi.
\newblock {HYSTERY}: a hybrid scheduling and mapping approach to optimize
  temperature, energy consumption and lifetime reliability of heterogeneous
  multiprocessor systems.
\newblock {\em The Journal of Supercomputing}, 74(5):2213--2238, January 2018.

\bibitem{MEJ}
Athena Abdi and Hamid~R. Zarandi.
\newblock A meta heuristic-based task scheduling and mapping method to optimize
  main design challenges of heterogeneous multiprocessor embedded systems.
\newblock {\em Microelectronics Journal}, 87:1--11, 2019.

\bibitem{spea_2}
Athena Abdi and Hamid~R. Zarandi.
\newblock Improving the reliability of multicore embedded systems through an
  evolutionary-based task scheduling approach.
\newblock In {\em 2021 29th Iranian Conference on Electrical Engineering
  (ICEE)}, pages 556--561, 2021.

\bibitem{akbari_2017_GA}
Mehdi Akbari, Hassan Rashidi, and Sasan~H Alizadeh.
\newblock An enhanced genetic algorithm with new operators for task scheduling
  in heterogeneous computing systems.
\newblock {\em Engineering Applications of Artificial Intelligence}, 61:35--46,
  2017.

\bibitem{ammar_2016performance}
Manel Ammar, Mouna Baklouti, and Mohamed Abid.
\newblock The performance-energy tradeoff in embedded systems design: A survey
  of existing design space exploration tools and trends.
\newblock {\em International Journal of Computer Science and Information
  Security}, 14(5):381, 2016.

\bibitem{power_2016}
Gayathri Ananthanarayanan, Smruti~R. Sarangi, and M.~Balakrishnan.
\newblock Leakage power aware task assignment algorithms for multicore
  platforms.
\newblock In {\em 2016 IEEE Computer Society Annual Symposium on VLSI
  (ISVLSI)}, pages 607--612, 2016.

\bibitem{AsadiEydivand2015}
Mitra Asadi-Eydivand, Mohammad~Mehdi Ebadzadeh, Mehran Solati-Hashjin,
  Christian Darlot, and Noor Azuan~Abu Osman.
\newblock Cerebellum-inspired neural network solution of the inverse kinematics
  problem.
\newblock {\em Biol. Cybern.}, 109(6):561--574, 2015.

\bibitem{ashrafi2020it2}
Mohammad Ashrafi, Dilip~K. Prasad, and Chai Quek.
\newblock It2-gsetsk: An evolving interval type-ii tsk fuzzy neural system for
  online modeling of noisy data.
\newblock {\em Neurocomputing}, 407:1--11, 2020.

\bibitem{assayad_2013tradeoff}
Ismail Assayad, Alain Girault, and Hamoudi Kalla.
\newblock Tradeoff exploration between reliability, power consumption, and
  execution time for embedded systems.
\newblock {\em International Journal on Software Tools for Technology
  Transfer}, 15(3):229--245, 2013.

\bibitem{BAKLOUTI2018}
N.~Baklouti, A.~Abraham, and A.M. Alimi.
\newblock A beta basis function interval type-2 fuzzy neural network for time
  series applications.
\newblock {\em Engineering Applications of Artificial Intelligence},
  71:259--274, 2018.

\bibitem{chantem_2011}
Thidapat Chantem, X.~Sharon Hu, and Robert~P. Dick.
\newblock Temperature-aware scheduling and assignment for hard real-time
  applications on mpsocs.
\newblock {\em IEEE Transactions on Very Large Scale Integration (VLSI)
  Systems}, 19(10):1884--1897, 2011.

\bibitem{CHENG_temp_2021}
Long Cheng, Kai Huang, Liang Mi, Gang Chen, Alois Knoll, and Xiaoqin Zhang.
\newblock Peak temperature analysis and optimization for pipelined hard
  real-time systems.
\newblock {\em Information Sciences}, 575:666--697, 2021.

\bibitem{crc_thermal}
Raj~P Chhabra.
\newblock {\em CRC handbook of thermal engineering}.
\newblock CRC press, 2017.

\bibitem{JEP122H}
Joint Electron Device~Engineering Council.
\newblock Failure mechanisms and models for semiconductor devices.
\newblock Technical Report JEP122H, 2016.

\bibitem{Das15}
Ankit~Kumar Das, Kartick Subramanian, and Suresh Sundaram.
\newblock An evolving interval type-2 neurofuzzy inference system and its
  metacognitive sequential learning algorithm.
\newblock {\em {IEEE} Trans. Fuzzy Syst.}, 23(6):2080--2093, 2015.

\bibitem{das_2014combined}
Anup Das, Akash Kumar, Bharadwaj Veeravalli, Cristiana Bolchini, and Antonio
  Miele.
\newblock Combined dvfs and mapping exploration for lifetime and soft-error
  susceptibility improvement in mpsocs.
\newblock In {\em 2014 Design, Automation \& Test in Europe Conference \&
  Exhibition (DATE)}, pages 1--6. IEEE, 2014.

\bibitem{das_2018literature}
Anup~Kumar Das, Akash Kumar, Bharadwaj Veeravalli, and Francky Catthoor.
\newblock Literature survey on system-level optimizations techniques.
\newblock In {\em Reliable and Energy Efficient Streaming Multiprocessor
  Systems}, pages 33--44. Springer International Publishing, November 2017.

\bibitem{de2020fuzzy}
Paulo~Vitor de~Campos~Souza.
\newblock Fuzzy neural networks and neuro-fuzzy networks: A review the main
  techniques and applications used in the literature.
\newblock {\em Applied Soft Computing}, page 106275, 2020.

\bibitem{nsga}
K.~Deb, A.~Pratap, S.~Agarwal, and T.~Meyarivan.
\newblock A fast and elitist multiobjective genetic algorithm: Nsga-ii.
\newblock {\em IEEE Transactions on Evolutionary Computation}, 6(2):182--197,
  2002.

\bibitem{Ebadzadeh15}
Mohammad~Mehdi Ebadzadeh and Armin Salimi{-}Badr.
\newblock {CFNN:} correlated fuzzy neural network.
\newblock {\em Neurocomputing}, 148:430--444, 2015.

\bibitem{Ebadzadeh2017}
Mohammad~Mehdi Ebadzadeh and Armin Salimi-Badr.
\newblock {IC-FNN}: a novel fuzzy neural network with interpretable, intuitive,
  and correlated-contours fuzzy rules for function approximation.
\newblock {\em {IEEE} Trans. Fuzzy Syst.}, 26(3):1288--1302, 2018.

\bibitem{Ekhtiyari2019}
Zohreh Ekhtiyari, Vahidreza Moghaddas, and Hakem Beitollahi.
\newblock A temperature-aware and energy-efficient fuzzy technique to schedule
  tasks in heterogeneous {MPSoC} systems.
\newblock {\em The Journal of Supercomputing}, 75(8):5398--5419, March 2019.

\bibitem{ferrandi2010ant}
Fabrizio Ferrandi, Pier~Luca Lanzi, Christian Pilato, Donatella Sciuto, and
  Antonino Tumeo.
\newblock Ant colony heuristic for mapping and scheduling tasks and
  communications on heterogeneous embedded systems.
\newblock {\em IEEE Transactions on Computer-Aided Design of Integrated
  Circuits and Systems}, 29(6):911--924, 2010.

\bibitem{franco_2014reliability}
Jacopo Franco, Ben Kaczer, and Guido Groeseneken.
\newblock {\em Reliability of high mobility SiGe channel MOSFETs for future
  CMOS applications}.
\newblock Springer, 2014.

\bibitem{WU2021498}
Chu ge~Wu, Wei Li, Ling Wang, and Albert~Y. Zomaya.
\newblock An evolutionary fuzzy scheduler for multi-objective resource
  allocation in fog computing.
\newblock {\em Future Generation Computer Systems}, 117:498--509, 2021.

\bibitem{alain_2009}
Alain Girault and Hamoudi Kalla.
\newblock A novel bicriteria scheduling heuristics providing a guaranteed
  global system failure rate.
\newblock {\em IEEE Transactions on Dependable and Secure Computing},
  6(4):241--254, 2009.

\bibitem{power_2008}
Lee~Kee Goh, Bharadwaj Veeravalli, and Sivakumar Viswanathan.
\newblock Design of fast and efficient energy-aware gradient-based scheduling
  algorithms heterogeneous embedded multiprocessor systems.
\newblock {\em IEEE Transactions on Parallel and Distributed Systems},
  20(1):1--12, 2009.

\bibitem{henkel_2021dependable}
J{\"o}rg Henkel and Nikil Dutt.
\newblock {\em Dependable Embedded Systems}.
\newblock Springer Nature, 2021.

\bibitem{multiprocessor_book}
Michael H{\"u}bner and J{\"u}rgen Becker.
\newblock {\em Multiprocessor system-on-chip: hardware design and tool
  integration}.
\newblock Springer Science \& Business Media, 2010.

\bibitem{ANFIS}
J.-S.R. Jang.
\newblock {ANFIS}: adaptive-network-based fuzzy inference system.
\newblock {\em {IEEE} Trans. Syst., Man, Cybern.}, 23(3):665--685, 1993.

\bibitem{mr_cross_2018}
Masoomeh Karami, Athena Abdi, and Hamid~R Zarandi.
\newblock A cross-layer aging-aware task scheduling approach for multiprocessor
  embedded systems.
\newblock {\em Microelectronics Reliability}, 85:190--197, 2018.

\bibitem{DENFIS}
N.K. Kasabov and Qun Song.
\newblock {DENFIS}: dynamic evolving neural-fuzzy inference system and its
  application for time-series prediction.
\newblock {\em {IEEE} Trans. Fuzzy Syst.}, 10(2):144--154, 2002.

\bibitem{Ebadzadeh09}
Omid Khayat, Mohammad~Mehdi Ebadzadeh, Hamid~Reza Shahdoosti, Ramin Rajaei, and
  Iman Khajehnasiri.
\newblock A novel hybrid algorithm for creating self-organizing fuzzy neural
  networks.
\newblock {\em Neurocomputing}, 73:517--524, 2009.

\bibitem{lee_2021thermal}
Youngmoon Lee.
\newblock Thermal-aware design and management of embedded real-time systems.
\newblock In {\em 2021 Design, Automation \& Test in Europe Conference \&
  Exhibition (DATE)}, pages 1252--1255. IEEE, 2021.

\bibitem{LUO2019}
Chao Luo, Chenhao Tan, Xingyuan Wang, and Yuanjie Zheng.
\newblock An evolving recurrent interval type-2 intuitionistic fuzzy neural
  network for online learning and time series prediction.
\newblock {\em Applied Soft Computing}, 78:150--163, 2019.

\bibitem{power_2021}
Hao Lv, Yong Guo, and Liu Yang.
\newblock Study of energy-efficient scheduling in multi-core systems in dynamic
  voltage and frequency adjustment.
\newblock In {\em Journal of Physics: Conference Series}, volume 2033, page
  012143. IOP Publishing, 2021.

\bibitem{chantem_hotspot}
Yue Ma, Thidapat Chantem, Robert~P. Dick, and Xiaobo~Sharon Hu.
\newblock Improving system-level lifetime reliability of multicore soft
  real-time systems.
\newblock {\em IEEE Transactions on Very Large Scale Integration (VLSI)
  Systems}, 25(6):1895--1905, 2017.

\bibitem{chantem_2021}
Yue Ma, Junlong Zhou, Thidapat Chantem, Robert~P Dick, and X~Sharon Hu.
\newblock Resource management for improving overall reliability of
  multi-processor systems-on-chip.
\newblock {\em Dependable Embedded Systems}, page 233, 2021.

\bibitem{chantem_2021_TCAD}
Yue Ma, Junlong Zhou, Thidapat Chantem, Robert~P. Dick, Shige Wang, and
  Xiaobo~Sharon Hu.
\newblock Online resource management for improving reliability of real-time
  systems on “big–little” type mpsocs.
\newblock {\em IEEE Transactions on Computer-Aided Design of Integrated
  Circuits and Systems}, 39(1):88--100, 2020.

\bibitem{Malek11}
Hamed Malek, Mohammad~Mehdi Ebadzadeh, and Mohammad Rahmati.
\newblock Three new fuzzy neural networks learning algorithms based on
  clustering, training error and genetic algorithm.
\newblock {\em Appl. Intell.}, 37(2):280--289, 2012.

\bibitem{marwedel_2021embedded}
Peter Marwedel.
\newblock {\em Embedded system design: embedded systems foundations of
  cyber-physical systems, and the internet of things}.
\newblock Springer Nature, 2021.

\bibitem{mendel2017uncertain}
Jerry~M Mendel.
\newblock Uncertain rule-based fuzzy systems.
\newblock {\em Introduction and new directions}, page 684, 2017.

\bibitem{tgff}
David Rhodes, Robert Dick, and Keith Vallerio.
\newblock Task graphs for free.
\newblock \url{http://ziyang.eecs.umich.edu/~dickrp/tgff}.
\newblock Accessed: 2022-02-09.

\bibitem{SOFMLS}
J.~de~Jesus Rubio.
\newblock {SOFMLS}: Online self-organizing fuzzy modified least-squares
  network.
\newblock {\em {IEEE} Trans. Fuzzy Syst.}, 17(6):1296--1309, 2009.

\bibitem{SALIMIBADR2022108258}
Armin Salimi-Badr.
\newblock It2cfnn: An interval type-2 correlation-aware fuzzy neural network to
  construct non-separable fuzzy rules with uncertain and adaptive shapes for
  nonlinear function approximation.
\newblock {\em Applied Soft Computing}, 115:108258, 2022.

\bibitem{Salimi-Badr2017}
{Armin} Salimi-Badr, {Mohammad M.} Ebadzadeh, and {Christian} Darlot.
\newblock Fuzzy neuronal model of motor control inspired by cerebellar pathways
  to online and gradually learn inverse biomechanical functions in the presence
  of delay.
\newblock {\em Biol. Cybern.}, 111(5-6):421--438, 2017.

\bibitem{SALIMIBADR2022139}
Armin Salimi-Badr and Mohammad~Mehdi Ebadzadeh.
\newblock A novel learning algorithm based on computing the rules’ desired
  outputs of a tsk fuzzy neural network with non-separable fuzzy rules.
\newblock {\em Neurocomputing}, 470:139--153, 2022.

\bibitem{SalimiBadr2022}
Armin Salimi-Badr and Mohammad~Mehdi Ebadzadeh.
\newblock A novel self-organizing fuzzy neural network to learn and mimic
  habitual sequential tasks.
\newblock {\em {IEEE} Transactions on Cybernetics}, 52(1):323--332, January
  2022.

\bibitem{scheffer2018eda}
Louis Scheffer, Luciano Lavagno, and Grant Martin.
\newblock {\em EDA for IC system design, verification, and testing}.
\newblock CRC press, 2018.

\bibitem{sheikh_2016sixteen}
Hafiz~Fahad Sheikh and Ishfaq Ahmad.
\newblock Sixteen heuristics for joint optimization of performance, energy, and
  temperature in allocating tasks to multi-cores.
\newblock {\em ACM Transactions on Parallel Computing (TOPC)}, 3(2):1--29,
  2016.

\bibitem{softcom_ant_2018}
G~Umarani Srikanth and R~Geetha.
\newblock Task scheduling using ant colony optimization in multicore
  architectures: a survey.
\newblock {\em Soft Computing}, 22(15):5179--5196, 2018.

\bibitem{srinivasan2004case}
Jayanth Srinivasan, Sarita~V Adve, Pradip Bose, and Jude~A Rivers.
\newblock The case for lifetime reliability-aware microprocessors.
\newblock {\em ACM SIGARCH Computer Architecture News}, 32(2):276, 2004.

\bibitem{nbti_2021bias}
Yen Tran, Toshihiro Nomura, Mohamed~Salim Cherchali, Claire Tassin, Yann Deval,
  and Cristell Maneux.
\newblock Bias temperature instability characterization and modeling for 0.18
  um cmos under extreme thermal stress conditions.
\newblock In {\em SMACD/PRIME 2021; International Conference on SMACD and 16th
  Conference on PRIME}, pages 1--4. VDE, 2021.

\bibitem{failure_2009esd}
Steven~H Voldman.
\newblock {\em ESD: failure mechanisms and models}.
\newblock John Wiley \& Sons, 2009.

\bibitem{cortex_2011}
Wei Wang and Tanima Dey.
\newblock A survey on arm cortex a processors.
\newblock {\em Retrieved March}, 2011.

\bibitem{wolf2009multiprocessor}
Wayne Wolf.
\newblock Multiprocessor system-on-chip technology.
\newblock {\em IEEE Signal Processing Magazine}, 26(6):50--54, 2009.

\bibitem{DFNN}
Shiqian Wu and Meng~Joo Er.
\newblock Dynamic fuzzy neural networks-a novel approach to function
  approximation.
\newblock {\em {IEEE} Trans. Syst., Man, Cybern. B}, 30(2):358--364, 2000.

\bibitem{GDFNN}
Shiqian Wu, Meng~Joo Er, and Yang Gao.
\newblock A fast approach for automatic generation of fuzzy rules by
  generalized dynamic fuzzy neural networks.
\newblock {\em {IEEE} Trans. Fuzzy Syst.}, 9(4):578--594, 2001.

\bibitem{xu2014_genetic}
Yuming Xu, Kenli Li, Jingtong Hu, and Keqin Li.
\newblock A genetic algorithm for task scheduling on heterogeneous computing
  systems using multiple priority queues.
\newblock {\em Information Sciences}, 270:255--287, 2014.

\bibitem{thiele_2013}
Hoeseok Yang, Iuliana Bacivarov, Devendra Rai, Jian-Jia Chen, and Lothar
  Thiele.
\newblock Real-time worst-case temperature analysis with temperature-dependent
  parameters.
\newblock {\em Real-Time Systems}, 49(6):730--762, 2013.

\bibitem{yoo_2018low}
Hoi-Jun Yoo, Kangmin Lee, and Jun~Kyoung Kim.
\newblock {\em Low-power noc for high-performance soc design}.
\newblock CRC press, 2018.

\bibitem{hybrid_2021}
Hassan Youness, Aly Omar, and Mohamed Moness.
\newblock An optimized weighted average makespan in fault-tolerant
  heterogeneous mpsocs.
\newblock {\em IEEE Transactions on Parallel and Distributed Systems},
  32(8):1933--1946, 2021.

\bibitem{Zadeh75}
L.~A. Zadeh.
\newblock The concept of a linguistic variable and its application to
  approximate reasoning.
\newblock {\em Journal of Information Science}, page 199, 1975.

\bibitem{zhou_power}
Dakai Zhu, R.~Melhem, and D.~Mosse.
\newblock The effects of energy management on reliability in real-time embedded
  systems.
\newblock In {\em IEEE/ACM International Conference on Computer Aided Design,
  2004. ICCAD-2004.}, pages 35--40, 2004.

\end{thebibliography}





\end{document}